\documentclass[preprint,12pt]{elsarticle}

\usepackage{amssymb}
\usepackage{amsthm}
\usepackage[dvipsnames]{xcolor}
\usepackage{tabularx}
\usepackage{amsmath}
\usepackage{diagbox}
\usepackage{soul} 

\usepackage[utf8]{inputenc}

\usepackage{amssymb}
\usepackage{amsmath,mathrsfs}
\usepackage{color}

\usepackage{extarrows}

\usepackage{verbatim}

\usepackage{lineno}

\usepackage{siunitx}

\usepackage{stfloats}

\usepackage{bm}

\usepackage{graphicx}

\usepackage{caption}
\usepackage{subcaption}

\usepackage{multicol}

\usepackage{multirow}

\usepackage{mhchem}

\usepackage[toc,page]{appendix}

\usepackage{bm}

\usepackage{float}

\usepackage{threeparttable}

\usepackage{sectsty}
\usepackage[toc,page]{appendix}
\usepackage{algorithm}     
\usepackage{algpseudocode} 
\begin{document}
\begin{frontmatter}

\title{A thermo-mechanical phase-field fracture model: \\ application to hot cracking simulations in additive manufacturing}

\author{Hui Ruan$^{1,*}$, Shahed Rezaei$^{1,*}$, Yangyiwei Yang$^{1}$, Dietmar Gross$^{2}$,  Bai-Xiang Xu$^{1,*}$} 
\address{$^1$Mechanics of Functional Materials Division, Institute of Materials Science,\\ Technische Universität Darmstadt, Darmstadt 64287, Germany}
\address{$^2$Division of Solid Mechanics,\\ Technische Universität Darmstadt, Darmstadt 64287, Germany}
\address{$^*$ corresponding authors: hui.ruan@tu-darmstadt.de, s.rezaei@mfm.tu-darmstadt.de, xu@mfm.tu-darmstadt.de}

\begin{abstract}
Thermal fracture is prevalent in many engineering problems and is one of the most devastating defects in metal additive manufacturing. Due to the interactive underlying physics involved, the computational simulation of such a process is challenging. In this work, we propose a thermo-mechanical phase-field fracture model, which is based on a thermodynamically consistent derivation. The influence of different coupling terms such as damage-informed thermomechanics and heat conduction and temperature-dependent fracture properties, as well as different phase-field fracture formulations, are discussed. 
Finally, the model is implemented in the finite element method and applied to simulate the hot cracking in additive manufacturing. Thereby not only the thermal strain but also the solidification shrinkage are considered. As for thermal history, various predicted thermal profiles, including analytical solution and numerical thermal temperature profile around the melting pool, are regarded, whereas the latter includes the influence of different process parameters. The studies reveal that the solidification shrinkage strain takes a dominant role in the formation of the circumferential crack, while the temperature gradient is mostly responsible for the central crack. Process parameter study demonstrates further that a higher laser power and slower scanning speed are favorable for keyhole mode hot cracking while a lower laser power and quicker scanning speed tend to form the conduction mode cracking. The numerical predictions of the hot cracking patterns are in good agreement with similar experimental observations, showing the capability of the model for further studies.

\end{abstract}






\end{frontmatter}

\section{Introduction}
\label{sec:sample1}
Fracture is a prominent issue for many structures and components in the engineering field. In various practical applications, fracture is coupled with other involved physics which in return severely influence the damage progression inside the material. As an example, when a piece of solid material is subjected to a sharp temperature variation, the thermal stress induced by the non-uniform thermal expansion results in breakage once it exceeds the material fracture strength. The latter phenomenon is known as thermal fracture or thermal shock. Specifically, for metal additive manufacturing (AM), thermal fracture plays an important role. Here, the subsequent layers of powders or wires are applied on top of the previous layers until the whole component is manufactured. Therefore, the material layers undergo complex thermal cycling history, where a non-uniform thermal field is formed and gives rise to thermal stress. In common practice, the thermal fracture happens, also known as hot cracking in AM.

\subsection{Hot cracking in additive manufacturing}
\begin{figure}[H]
    \centering
    \includegraphics[width=0.8\textwidth]{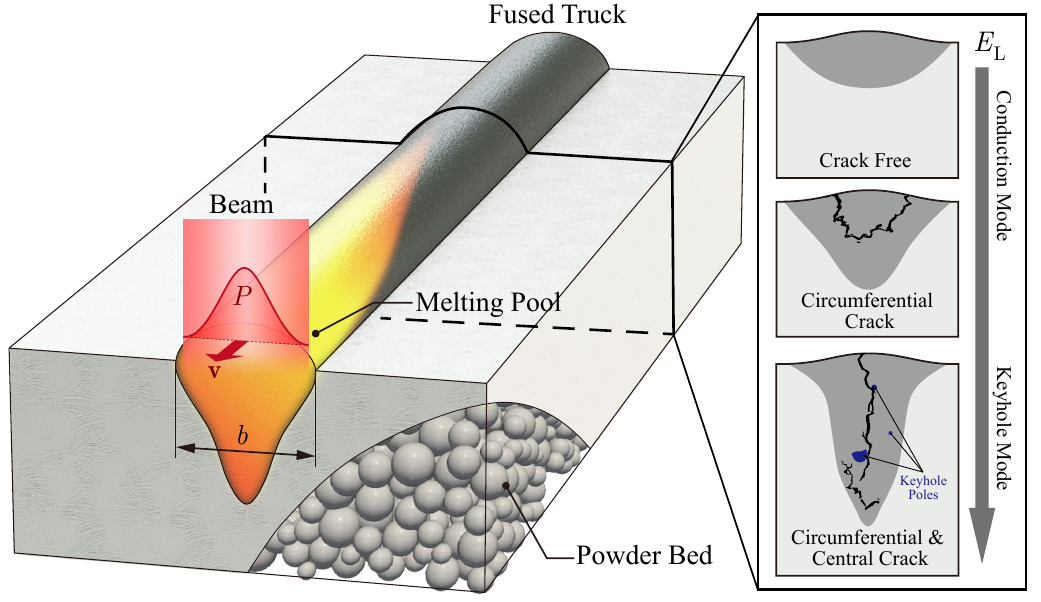}
    \caption{Schematic of PBF process and hot cracking phenomenon. Different hot cracking patterns in the melting pool for conduction mode and keyhole mode hot cracking with the increase of energy density. }
    \label{fig:fig_scheme}
\end{figure}

There are many defects in the process of AM, such as porosity, inclusion, unmelting, and residual distortion \cite{khairallah2020controlling, liu2019additive, zhang2017defect}. Hot cracking is one of the most common and devastating defects that hinder the widespread application of AM in the engineering field. As shown in Fig. \ref{fig:fig_scheme}, in the process of AM, specifically Powder Bed Fusion (PBF), different hot cracking patterns are observed in the cross-section of single track samples. Typically for the conduction mode AM process, the hot cracking shows as circumferential crack, while for the keyhole mode AM process, it shows as a combination of circumferential crack and central crack. The two different patterns are referred to as the conduction mode hot cracking and keyhole mode hot cracking, respectively.

Hot cracking is the formation of cracks in and around the melt pool during the AM. The low melting compositions solidify at the gap of dendrite and occur as intercrystalline and/or interdendritic cracks \cite{wang2019cracking, chauvet2018hot, lu2020hot}. This process is the result of the competition between the mechanical driving forces (mainly from the thermal stress) and the material’s intrinsic resistance to cracking \cite{chen2016dendritic}. It is a dynamic process covering multi-physics at different time scales and length scales, which makes the investigation of this phenomenon extremely challenging. Some works have been reported concerning microstructural characterization and hot cracking susceptibility in the different AM processes. Recently, Wu et al. \cite{wu2022hot} established the relationship between the microstructural evolution and hot cracking susceptibility to reveal the hot craking mechanism by three-dimensional X-ray micro-tomography technique, see also \cite{xin2021hot}. Chauvet et al.  \cite{chauvet2018hot} elucidated the mechanism and the origin of cracking in a non-weldable Ni-based superalloy fabricated by selective electron beam melting (S-EBM) and concluded that the presence of liquid films is required and the hot cracking susceptibility depends on the grain boundary misorientation. Stopyra et al. \cite{stopyra2020laser} aimed at finding the process window for the laser powder bed fusion (LPBF) manufacturing of defect-free components of AA7075 alloy, and found that the solidification cracks are formed by the liquid film rupture mode. Lu et al. \cite{lu2020hot} obtained the same results with directed energy deposition. The effect of thermal gradient on the hot cracking tendency is also studied. In the work of Chen et al. \cite{chen2016dendritic}, the effect of thermal gradient on the grain boundary misorientation and hot cracking tendency during the laser additive manufacturing of IN718. The result shows that highly ordered dendrites were established and liquation cracking tendency was reduced under improved base cooling.

There are also some simulation works conducted from the perspective of microstructure characterization to gain a better understanding of the process \cite{debroy2021metallurgy, bayat2021review, wei2021mechanistic}. Gao et al. \cite{gao2018hot} studied the hot cracking in laser welding with a multi-scale modeling approach, where the thermal field is calculated with the finite element method and microstructural evolution by phase-field modeling. Vrancken et al. \cite{vrancken2020analysis} combined thermo-mechanical simulations with in situ high-speed videos of microcracking in single laser-melted tracks and found microcracking occurs in a narrow temperature interval. A multi-scale model was developed by Nie et al. \cite{nie2014numerical} by combining the finite element method and stochastic analysis.

While extensive experimental studies on additive manufactured microstructures and hot cracking have been performed, the mechanism is still not fully understood. The latter point limits the process optimization to avoid cracking and to improve product quality. An important task, therefore, is to derive physical models and carry out reliable numerical simulations. Surprisingly, the numerical model of hot cracking is very limited, particularly not in multiphysics scenario.

\subsection{Phase-field fracture model}
Computational modeling of thermal fracture has been studied extensively with different approaches.  In particular, phase-field modeling of fracture in solids has received extensive theoretical and computational attention. In the framework of the phase-field method, the sharp interface is replaced by a continuous field variable, i.e. order parameter to differentiate multiple phases smoothly. Specifically, for the phase-field fracture (PFF) model, the non-smooth crack is regularized by the diffusive crack with an auxiliary field variable $d$ in the range of $0$ to $1$, with $0$ denoting unbroken state and $1$ fully broken state. The regularization is governed by the length-scale parameter $\ell_{0}$ which controls the width of the diffusive zone. The diffusive crack topology approaches the sharp crack topology when $\ell_{0}$ goes towards 0 according to the $\Gamma$-convergence \cite{braides1998approximation}. Therefore, PFF describes the crack interface continuously and no additional tracking of the surface is needed. Meanwhile, based on the variational framework, it does not need an ad-hoc failure criterion and the framework makes it easy to incorporate other physics (e.g. thermal field) as well.

The phase-field modeling of brittle fracture originated from the work by Francfort and Marigo \cite{francfort1998revisiting}. They proposed a phase-field fracture model from the variational approach to brittle fracture of reformulating Griffith's energy criterion \cite{griffith1921vi}. It aims to get the displacement field and fracture field simultaneously by minimizing the total potential energy as the sum of the fracture energy and elastic energy of the crack system. Karma et al. \cite{karma2001phase} proposed conceptually similar phase-field approach to brittle fracture based on the classical Ginzburg-Landau evolution equation. The numerical implementation of the model of brittle fracture developed in \cite{francfort1998revisiting} and \cite{bourdin2000numerical}, see also \cite{bourdin2008variational, amor2009regularized}.
PFF has the flexibility to simulate crack initiation, propagation, merging as well as branching, and thereby is extended to various fracture problems, such as ductile fracture \cite{hu2021variational, ambati2015phase}, dynamic fracture \cite{borden2012phase, geelen2019phase}, hydraulic fracture \cite{miehe2016phase, ULLOA2022115084}, viscoelastic fracture \cite{yin2020fracture}, anisotropic fracture \cite{khosravani2022fracture, REZAEI2021} and multiphysics fracture problems \cite{xu2016phase, xu2010phase}.

PFF model is also applied to study thermal fracture. Miehe et al. \cite{miehe2015phase} extended the PFF toward coupled thermo-mechanical and multi-physics problems at large strain. Li et al. \cite{li2021multiphysics} presented phase-field modeling of quasi-static cracking in urania ceramic under multiphysics including neutron radiation and high-temperature. Different PFF models apart from the standard PFF model are also utilized. For example, Tangella et al. \cite{tangella2021hybrid} presented numerical modeling of the thermo-elastic fracture using a hybrid phase-field method.
Mandal et al. \cite{mandal2021fracture} extended the phase-field regularized cohesive zone model (PF-CZM) into multiphysics and presented a monolithic BFGS algorithm to solve thermo-elastic fracture.
Some works also studied thermal fracture of single crystal and polycrystalline materials. Razwan et al. \cite{rezwan2021modeling} investigated brittle fracture of polycrystalline materials due to thermal stress arising from anisotropic thermal expansion. Zhang et al. \cite{zhang2021phase} proposed a phase-field fracture model that includes single crystal anisotropy in both the elastic constants and the fracture energy. 
Li et al. \cite{li2021phase} developed a phase-field approach to model hydro-thermally induced crack propagation in thermo-poroelastic media.
Badnava et al. \cite{badnava2018h} investigated thermo-mechanical induced cracks using a phase-field model in 2D and 3D continua in homogeneous and heterogeneous materials. The influence of crack on the thermal field was also studied, specifically the heat conduction.
Svolos et al. \cite{svolos2020thermal} proposed a thermal-conductivity degradation function derived from a novel micromechanics analytical approach using spherical harmonics, and showed that the thermal conductivity across cracks must be degraded to satisfy crack Neumann boundary conditions. Furthermore, they proposed a new anisotropic approach~\cite{svolos2021anisotropic}  in which thermal conductivity, which depends on the phase-field gradient, is degraded solely across the crack.

Despite the previously mentioned progress, the thermodynamic consistency of the fully coupled phase-field model for thermal fracture is not investigated properly. One of our first goals in this work is to develop a consistent framework for the phase-field model of thermal fracture, in which various coupling aspects between different fields are included. In this model, the temperature field interacts with the displacement field and the crack field. The coupling effects like damage-informed thermoelasticity and heat conduction, and temperature-dependent fracture properties are interactively considered. Particularly, the degradation of elastic energy and thermal conductivity due to cracking and the temperature-dependent fracture toughness are studied to recapture the heat transfer behavior and the crack pattern more accurately. Furthermore, to the authors' best knowledge such a method is not yet applied to studies on hot cracking in AM process which belongs to our next main goal in this contribution.

Based on the reviewed literature, this paper aims to cover some shortcomings related to thermo-fracture modeling as well as the application of the PFF model in AM process. In Section 2, the thermodynamically consistent phase-field model for thermo-elastic brittle fracture is presented, which is derived from the basic principles of thermodynamics. In Section 3, the weak form and numerical discretization of the problem for implementation with the finite element method are provided. In Section 4, several numerical examples are then presented. The model is validated by the canonical single edge notched tension test and further quenching test. Subsequently, the proposed model is applied to study the different hot cracking patterns in AM process. The influence of different process parameters on the hot cracking patterns is also investigated. Finally, the conclusions and outlook of the current study are presented.

\section{A thermo-mechanical phase-field brittle fracture model}\label{sec:sample1}
The current work focuses on the thermo-mechanical coupling and its impact on hot cracking during PBF. To single out the phenomenon, we assume first small deformation, elastostatics and the brittle fracture behavior. Such assumptions are applicable for the brittle materials e.g.~ceramics or brittle glass. For metallic materials, plasticity and ductile fracture should be addressed. However, it should be noted that in any case they share some common thermo-mechanical coupling mechanisms, which is the objective of this work. Soon after these mechanisms are understood in the linear and brittle scenario, the extension of this framework for the nonlinear deformation and for the elastoplastic ductile fracture can be expected in the next steps.
For thermo-elastic coupled brittle fracture, the primary field variables consist of the displacement field $\boldsymbol u (\boldsymbol x,t)$, the damage field $d(\boldsymbol x,t)$ and the temperature field $T(\boldsymbol x,t)$.


\subsection{Energy dissipation inequality}\label{sec:da}
The second law of thermodynamics, which is expressed in the form of local Clausius-Duhem inequality is utilized here to derive the thermodynamically consistent constitutive laws of the model. The detailed derivation is provided in the appendix A. For this thermoelastic coupled problem, the local dissipated energy $\mathcal{D}$ considering the power produced by the micro and macro forces, is given as
\begin{equation}\label{eq:Clausius-Duhem}
    \mathcal D = \boldsymbol\sigma: \dot{\boldsymbol\varepsilon}+K \dot{d}+\boldsymbol H \cdot \nabla \dot{d}-\rho(\dot{\psi}+\dot{T}\eta )-\dfrac{1}{T} \nabla T \cdot \boldsymbol{q} \geq 0.
\end{equation}


Here, $\boldsymbol \sigma$ is the Cauchy stress tensor, $\boldsymbol \varepsilon$ is the total strain, $\psi$ denotes the Helmholtz free energy, $\eta$ denotes the entropy, $\boldsymbol H$ is micro-traction on crack surfaces, $K$ is the internal micro-forces, and $\rho$ is the material density.

The Helmholtz free energy $\psi$ is decomposed into elastic energy $\psi_e$, fracture energy $\psi_c$ and thermal energy $\psi_T$ parts, as follows
\begin{equation}\label{eq:total_free_energy}
    \psi = \psi (\boldsymbol\varepsilon,  d, \nabla d,T) = \psi_{e}(\boldsymbol\varepsilon,d,T)+\psi_{c}(d,\nabla d,T)+\psi_{T}(T).
\end{equation}
The specific energy terms will be explained in the following subsections. For the above equation, the rate of the free energy change is given by
\begin{equation}\label{eq:free_energy_dot}
    \dot{\psi}=\dot{\psi}_{e}+\dot{\psi}_{c}+\dot{\psi}_{T},
\end{equation}
where for the rate of each component we have
\begin{equation}\label{eq:three_energy}
\left\{\begin{array}{ll}

\dot{\psi}_{e}=\dfrac{\partial \psi_{e}}{\partial \boldsymbol\varepsilon}: \dot{\boldsymbol\varepsilon}+\dfrac{\partial \psi_{e}}{\partial d} \dot{d}+ \dfrac{\partial \psi_{e}}{\partial T} \dot{T} \\ \\

\dot{\psi}_{c}=\dfrac{\partial \psi_{c}}{\partial d} \dot{d}+\dfrac{\partial \psi_{c}}{\partial(\nabla d)} \cdot \nabla \dot{d}+\dfrac{\partial \psi_{c}}{\partial T} \dot{T} \\ \\

\dot{\psi}_{T}=\dfrac{\partial \psi_{T}}{\partial T} \dot{T}

\end{array}\right.
\end{equation}
Substituting Eq.~\ref{eq:three_energy} into Eq.~\ref{eq:free_energy_dot} and regrouping terms, the total free energy rate reads:
\begin{equation}\label{eq:24}
    \dot{\psi}=\dfrac{\partial \psi_{e}}{\partial \boldsymbol\varepsilon}: \dot{\boldsymbol\varepsilon}+\left(\dfrac{\partial \psi_{e}}{\partial d}+\dfrac{\partial \psi_{c}}{\partial d}\right) \dot{d}+\dfrac{\partial \psi_{c}}{\partial(\nabla d)} \cdot \nabla \dot{d}+(\dfrac{\partial \psi_{e}}{\partial T}+\dfrac{\partial \psi_c}{\partial T} +\dfrac{\partial \psi_T}{\partial T})\dot{T}.
\end{equation}
Therefore, the Clausius-Duhem inequality (Eq.~\ref{eq:Clausius-Duhem}) is rewritten as

\begin{equation}\label{eq:Clausius-Duhem_new}
    \left(\boldsymbol \sigma-\rho \dfrac{\partial \psi}{\partial \boldsymbol \varepsilon}\right): \dot{\boldsymbol \varepsilon}+\left(K-\rho \dfrac{\partial \psi}{\partial d}\right) \dot{d}+\left(\boldsymbol{H}-\rho \dfrac{\partial \psi}{\partial(\nabla d)}\right) \cdot \nabla \dot{d} \\-\left(\rho \eta+\rho \dfrac{\partial \psi}{\partial T}\right) \dot{T}-\dfrac{1}{T} \nabla T \cdot \boldsymbol{q} \geq 0.
\end{equation}

Note that here elastostatic and quasi-static fracture are adopted. The inequality in Eq.~\ref{eq:Clausius-Duhem_new} must hold for any arbitrary thermodynamic processes. Hence, the coefficients of the dissipative terms are non-negative while the coefficients of the non-dissipative terms must vanish. Following the Coleman-Noll procedure \cite{coleman1974thermodynamics} for a thermodynamically consistent model the thermoelastic laws are

\begin{equation}\label{eq:TherForc}
    \left\{\begin{array}{ll}
\text{Elastic stress tensor:}&\boldsymbol\sigma=\rho \dfrac{\partial \psi}{\partial \boldsymbol\varepsilon}=\rho \dfrac{\partial \psi_{e}}{\partial \boldsymbol\varepsilon}  \\ \\
\text{Micro-traction equation:}&\boldsymbol{H}=\rho \dfrac{\partial \psi}{\partial(\nabla d)}=\rho \dfrac{\partial \psi_{c}}{\partial(\nabla d)} \\ \\
\text{Internal micro-force equation:}&K=\rho\dfrac{\partial\psi}{\partial d} =\rho \dfrac{\partial \psi_{e}}{\partial d}+\rho \dfrac{\partial \psi_{c}}{\partial d}\\ \\
\text{Entropy equation:}&\eta=-\dfrac{\partial \psi}{\partial T}

\end{array}\right.
\end{equation}
Following the above assumptions for the thermodynamics forces, the remaining part of the dissipation inequality reads:
\begin{equation}\label{eq:heat_ineq}
-\nabla T \cdot \boldsymbol{q} \geq 0.
\end{equation}
The above relation is also referred to as heat conduction inequality.


\subsection{Damage informed thermoelasticity}
The balance of the linear momentum equation in the tensorial notation and in the absence of the body force reads
\begin{equation}\label{eq:27}
    \nabla \cdot \boldsymbol{\sigma}=0,
\end{equation}
where $\nabla \cdot$ is the divergence operator. The total strain $\boldsymbol \varepsilon$ is additively decomposed into the elastic part and the thermal part:
\begin{equation}\label{eq:29}
    \boldsymbol{\varepsilon}=\boldsymbol{\varepsilon}_{e}+\boldsymbol{\varepsilon}_{t}=\nabla_{s} \boldsymbol{u}:=\dfrac{1}{2}\left(\nabla \boldsymbol{u}+\nabla^{T} \boldsymbol{u}\right),
\end{equation}
where $\boldsymbol{\varepsilon}_{e}$ denotes the elastic strain. The thermal strain $\boldsymbol{\varepsilon}_{t}$ follows a linear expansion law:
\begin{equation}\label{eq:30}
    \boldsymbol{\varepsilon}_{t}=\alpha_{t} ({T-T_0}) \boldsymbol I,
\end{equation}
where $\alpha_{t}$ is thermal expansion coefficient, $T_0$ is the initial temperature and $\boldsymbol I$ is the second-order identity tensor.
The total elastic free energy density for an undamaged body can be expressed as
\begin{equation}\label{eq:32}
    \psi_e=\dfrac{1}{2} \boldsymbol\varepsilon_{e}: \mathbb{C}^{e}: \boldsymbol\varepsilon_{e} =\dfrac{\lambda}{2}(\text{tr}~\varepsilon_{e})^{2}+\mu  \text{tr}(\varepsilon_{e}^{2}).
\end{equation}
The fourth-order elastic stiffness tensor is denoted by $\mathbb{C}^{e}$, and is expressed for isotropic elastic materials in terms of Lame constant $\lambda$ and $\mu$ as
\begin{equation}\label{eq:33}
    \mathbb{C}^{e}=\lambda \boldsymbol{I} \otimes \boldsymbol{I}+2 \mu \mathbb{I}_{S},
\end{equation}
where $\mathbb{I}_{S}$ is the symmetric fourth-order identity tensor, $\otimes$ denotes the dyadic product of two second order tensors. In the indicial notation, the symmetric fourth-order identity tensor is expressed $(\mathbb{I}_{S})_{ijkl} = \dfrac{1}{2} (\delta_{ik} \delta_{jl} + \delta_{il} \delta_{jk})$, where $\delta_{ij}$ is the Kronecker symbol, and the second-order identity tensor is defined as $I_{ij} = \delta_{ij}$.

To differentiate degradation in tension from compression, we additively decompose elastic strain energy density $\psi_e$ into a positive (tensile) part $\psi_e^+$ and a negative (compressive) part $\psi_e^-$:
\begin{equation}\label{eq:34}
    \psi_{e}=\psi_e ^{-}+f(d) \psi_e^{+}.
\end{equation}
Here the function $f(d)$ is the so-called degradation function, which has the following properties:
\begin{equation}\label{eq:degradation_function}
    f(d)\in [0,1], f(0)=1, f(1)=0; f'(d)<0, f'(1)=0.
\end{equation}
The choice of $f(d)$ will be discussed in detail in the following. For the decomposition method, the spectral decomposition of the strain tensor is utilized:
\begin{equation}
    \boldsymbol\varepsilon_e=\boldsymbol\varepsilon_{e}^++\boldsymbol\varepsilon_{e}^-, \quad 
\boldsymbol\varepsilon_{e}^{\pm}=\sum_{i=1}^{3} \left\langle\varepsilon_{i} \right\rangle ^{\pm} \boldsymbol{n}_{i} \otimes \boldsymbol{n}_{i}.
\end{equation}
Here, $\varepsilon_{i=1,2,3}$ are the principal strains and $\boldsymbol n_{i=1,2,3}$ denote the principal strain directions. The Macaulay bracket operator is defined as $\langle x\rangle=\left\{\begin{array}{ll}
x & \text { if } x \geq 0, \\
0 & \text { if } x<0.
\end{array}\right.$
Therefore, for the different parts of the elastic energy we have
\begin{equation}
    \psi_e ^{\pm}=\dfrac{\lambda}{2}({\left\langle \text{tr}~\boldsymbol \varepsilon_e\right\rangle^{\pm}})^{2}+\mu \text{tr}\left({\left\langle\boldsymbol\varepsilon_e\right\rangle^{\pm}}^{2}\right).
\end{equation}
In this formulation, only the tensile component contributes to fracture. Similarly, the degraded stress tensor can be derived as 

\begin{equation}
    \boldsymbol \sigma=\dfrac{\partial \psi_e}{\partial \boldsymbol \varepsilon}=f(d)\dfrac{\partial \psi_e^+}{\partial \boldsymbol \varepsilon} +\dfrac{\partial \psi_e^-}{\partial \boldsymbol \varepsilon}=f(d)\boldsymbol \sigma^+ +\boldsymbol \sigma^-,
\end{equation}
where
\begin{equation}
    \boldsymbol\sigma^{\pm}=\lambda{\left\langle \text{tr}~ \boldsymbol\varepsilon_e\right\rangle^{\pm}}+2\mu {\left\langle\boldsymbol\varepsilon_e\right\rangle^{\pm}}.
\end{equation}

In this paper, we explore also the applicability of the very promising cohesive phase-field (CPF) fracture model~\cite{lorentz2011convergence, wu2018length, geelen2019phase}, which characterizes itself by two main features: the threshold for damage initiation and insensitivity to the length scale $\ell_{0}$. Thereby the degradation function for elastic energy is defined as 
\begin{equation}
   f(d) = \dfrac{(1-d)^2}{(1-d)^2+a_1 d (1+a_2 d+a_3 d^2)}.
\end{equation}
In the above equation, $a_1=\dfrac{4~l_{ch}}{\pi \ell_{0}}$, $a_2=-\dfrac{1}{2}$ and $a_3=0$ are selected to represent the cohesive nature of fracture in the process zone. Furthermore, $\l_{ch}=\dfrac{EG_c}{\sigma_u^2}$ is the Irwin's length which measures the size of the fracture process zone. The smaller this length scale is, the more brittle the material behaves. The parameters $a_2$ and $a_3$ are the shape parameters and can be tuned to represent the different softening curves \cite{wu2018length}.

To prevent cracks from healing when $\psi_e^+$ decreases, the irreversibility condition is enforced. A history variable $\mathcal H$ is introduced, which must satisfy the Karush-Kuhn-Tucker (KTT) conditions:
\begin{equation}
    \psi_{e}^{+}-\mathcal{H} \leq 0, \quad \dot{\mathcal{H}} \geq 0, \quad \dot{\mathcal{H}}\left(\psi_{e}^{+}-\mathcal{H}\right)=0.
\end{equation}
Thus, the history variable can be written as
\begin{equation}\label{eq:history_variable}
    \mathcal{H}=\max _{t} (\psi_e^{+}(\boldsymbol\varepsilon_e,t), \psi_{th}).
\end{equation}
where the damage threshold $\psi_{th}$ is defined as 
\begin{equation}
    \psi_{th}=\dfrac{\sigma_u^2}{2E}.
\end{equation}
where $\sigma_u$ is the material strength and $E$ Young's modulus. Eq. \ref{eq:history_variable} implies that before the onset of any damage the material is characterized by an elastic domain, until the $\psi_{th}$ is reached.

\subsection{Temperature-dependent phase-field fracture model}
Consider a cracked solid $\Omega$ with an external boundary denoted by $\partial \Omega$ and a crack set $\Gamma$. Starting from Griffith's theory in the fracture mechanics, the total fracture energy is given by
\begin{equation}\label{eq:frac_energy}
    \Psi_{c}= \int_{\Gamma} G_{c}(T) \mathrm{~d} A =
    \int_\Omega G_c(T) \gamma (d, \nabla d) dV.
\end{equation}
In the above relation, we substitute the surface integral with a volumetric integral which yields an approximation of the fracture energy \cite{amor2009regularized, miehe2010phase}. Here, $\gamma$ is the crack surface density function and is defined as 
\begin{equation}
    \gamma(d, \nabla d)=\dfrac{1}{c_0}(\dfrac{1}{\ell_{0}}\omega(d)+\ell_{0}|\nabla d|^2), \quad c_0=4\int_0^1\sqrt{\omega(\beta)}d\beta.
\end{equation}
In the above equation, the geometric function $\omega(d)$ characterizes the homogeneous evolution of the phase-field crack, which has the properties
\begin{equation}
    \omega(d)\in [0,1], \omega(0)=0, \omega(1)=1; \omega'(d)>0.
\end{equation}
$\ell_{0}$ is the length scale parameter regularizing the sharp crack, which is related to the diffusive crack width, and finally, $c_0 > $ 0 is a scaling parameter (see also \cite{miehe2010phase, wu2018length}).

Different choices of the geometric function and degradation function results in different phase-field fracture models. The three most commonly used models are listed in Table~\ref{tab:PFmodel}. In the following session, the AT2 model, which is more often used so far than AT1, is chosen to compare with the CPF model. 
 \begin{table}[H]
      \centering
      \caption{Different phase-field fracture models.}
        \begin{tabular}{c c c c c c c c} \hline
        \diagbox{Features}{Models}& AT1 & AT2 & CPF  \\
        \hline
        $\omega( d )$    & $d$  & $d^2$    & $2d-d^2$    \\
        $c_0$ & $\dfrac{8}{3}$ & 2 & $\pi$\\
        $f(d)$    & $(1-d)^2$  & $(1-d)^2$    & $\dfrac{(1-d)^2}{(1-d)^2+a_1 d (1+a_2 d+a_3 d^2)}$   \\
        \hline
        \end{tabular}
        \label{tab:PFmodel}
\end{table}

For the AT2 model, the geometric function and degradation function takes the quadratic form, with which the predicted material strength shows a strong dependence on $\ell_{0}$ \cite{lorentz2017nonlocal}. In the work of Lorentz et al. \cite{lorentz2017nonlocal, lorentz2011convergence} and Wu et al. \cite{wu2018length}, a rational degradation function was proposed. It is shown that the material response for this formulation converges to the sharp interface behavior (cohesive zone) as $\ell_{0}$ decreases \cite{REZAEI2022108177}. The specific form of the length-scale-insensitive model is taken in this work. One advantage thereby lies in the fact that it allows to prescribe the ultimate strength $\sigma_u$, in addition to the fracture energy value $G_c$. 
As a result, the model takes the cohesive nature of the fracture into account and can produce numerical results which converge with respect to the internal length scale parameter. The latter point can be an interesting option for problems containing multiphysics fracture since the influence of the length scale on the results can be omitted.

For quasi-static fracture assuming that micro-inertia is negligible, the micro-force momentum balance equation is given by
\begin{equation}\label{eq:28}
    \nabla \cdot \boldsymbol{H}=K.
\end{equation}
where $\boldsymbol H$ and $K$ are defined in Eq. \ref{eq:TherForc}.
Employing the the micro-force balance equation and using the fracture free energy given in \ref{eq:frac_energy}, the phase-field governing equation reads:
\begin{equation}
    \dfrac{G_{c}}{c_0\ell_{0}} \omega'(d)- Y-\dfrac{2 G_{c}\ell_{0}}{c_0} \nabla^{2} d=0,
\end{equation}
where the driving force of the phase-field is defined as
\begin{equation}
    - Y=\dfrac{\partial \psi_e}{\partial d}=f'(d)\mathcal{H}.
\end{equation}

In thermo-mechanical problems, the temperature range can be considerable. Consequently, the variation of $G_c$ with the temperature near the crack tip cannot be ignored, which affects both the initiation of crack onset for a non-isothermal quasi-brittle fracture and the dynamics of crack propagation \cite{dittmann2019variational}. Thus the temperature-dependency of $G_c$ needs to be taken into account to capture the crack patterns more accurately. For brittle materials, the dependency of $G_c$ with temperature in this work is based on the description \cite{kitamura2008crack}. For quasi-brittle materials, the analytical relations of $G_c$ with temperature can be found in \cite{bayoumi1983temperature}. In this work, $G_c$ is considered temperature-dependent and takes the form as
\begin{equation}
    G_c={G_c}_0 [1-b_1 \dfrac{T-T_{ref}}{T_{max}}+b_2(\dfrac{T-T_{ref}}{T_{max}})^2].
    \label{Eq:Gc(T)}
\end{equation}
Here, $b_1$ and $b_2$ are constant model parameters, $T_{ref}$ and $T_{max}$ are the reference temperature and the maximum temperature, respectively, and ${G_c}_0$ is the value of $G_c$ at $T_{ref}$. In this work, $b_1=1.80$, $b_2=1.10$, $T_{ref}$ and $T_{max}$ are 300 K and 1000K, respectively. 
\subsection{Damaged informed heat conduction}
The energy balance equation, derived from Eq. \ref{eq:8} is written as
\begin{equation}\label{eq:energy_balance}
    \rho \dot{e}=\boldsymbol \sigma: \dot{\boldsymbol\varepsilon}+K \dot{d}+ H \cdot \nabla \dot{d}-\nabla\cdot \boldsymbol{q}+Q.
\end{equation}
Given the relation $e=\psi+T\eta$ one can write:
\begin{equation}\label{eq:entropy_energy}
    \dot e=\dot \psi+\dot T\eta+T\dot\eta.
\end{equation}
Considering $\psi=\psi(\boldsymbol \varepsilon,  d, \nabla d,T )$ and the thermodynamic relations obtained in Eq.~\ref{eq:TherForc}, we can obtain
\begin{equation}\label{eq:psidot}
    \dot{\psi}=\dfrac{1}{\rho}\left(\boldsymbol\sigma: \dot{\boldsymbol\varepsilon}+K \dot{d}+H \cdot \nabla \dot{d}-\rho\dot{T}\eta\right).
\end{equation}
Substituting Eq.~\ref{eq:psidot} into Eq.~\ref{eq:entropy_energy}, we have
\begin{equation}\label{eq:59}
\begin{aligned}
    \rho \dot e=\boldsymbol\sigma: \dot{\boldsymbol\varepsilon}+K \dot{d}+H \cdot \nabla \dot{d} +\rho T\dot\eta.
\end{aligned}
\end{equation}
Comparing Eqs.~\ref{eq:energy_balance} and \ref{eq:59}, one can conclude
\begin{equation}\label{eq:heat_eq1}
    \rho T\dot\eta=-\nabla\cdot \boldsymbol{q}+Q.
\end{equation}
Note that $\eta(\boldsymbol \varepsilon,  d, \nabla d,T )=-\dfrac{\partial \psi}{\partial T}$. Therefore for the specific entropy rate we have
\begin{equation}\label{eq:ent_rate}
\begin{aligned}
    \dot \eta &=
-\left(\dfrac{\partial^{2} \psi}{\partial T\partial \boldsymbol\varepsilon}: \dot{\boldsymbol\varepsilon}+\dfrac{\partial^{2} \psi}{\partial T\partial d } \dot d+\dfrac{\partial^{2} \psi}{\partial T\partial (\nabla d) } \nabla \dot d +\dfrac{\partial^{2} \psi}{\partial T^{2}} \dot{T}\right)\\
&=-\dfrac{1}{\rho}\left(\dfrac{\partial \boldsymbol \sigma}{\partial T}:\dot{\boldsymbol\varepsilon}+\dfrac{\partial K}{\partial T } \dot d+\dfrac{\partial H}{\partial T } \nabla \dot d-\rho\dfrac{\partial \eta}{\partial T} \dot{T}\right).
\end{aligned}
\end{equation}
Next, we consider the Fourier’s law by means of which the inequality relation in Eq.~\ref{eq:heat_ineq} is automatically satisfied:
\begin{equation} \label{eq:FL}
    \boldsymbol{q}=-k(d) \nabla T.
\end{equation}
Here ${k(d)}$ is the degraded thermal conductivity affected by the phase-field $d$ and it is expressed as
\begin{equation}\label{eq:thermal_condutivity}
    k(d)=g(d) k_0.
\end{equation}
In the above relation, $k_0$ is the thermal conductivity of the undamaged material, and $g(d)$ is a thermal degradation function which ensures that no heat flux exists across the crack.  Though there are other forms of thermal degradation proposed in the work of \cite{svolos2020thermal, svolos2021anisotropic}, an isotropic conductivity degradation $g(d)=(1-d)^2+\xi$ is adopted here, where $\xi$ is a small number for numerical and physical purposes.
Substituting Eq.~\ref{eq:ent_rate} into Eq.~\ref{eq:heat_eq1} we have
\begin{equation}\label{eq:64}
    - T\left(\dfrac{\partial \boldsymbol \sigma}{\partial T}:\dot{\boldsymbol\varepsilon}+\dfrac{\partial K}{\partial T } \dot d+\dfrac{\partial H}{\partial T } \nabla \dot d-\rho\dfrac{\partial \eta}{\partial T} \dot{T}\right) = k(d)\nabla^2T+Q.
\end{equation}
By introducing the specific heat defined as
\begin{equation}
    c = T\dfrac{\partial \eta}{\partial T},
\end{equation}
the complete form of the heat equation reads:
\begin{equation}\label{eq:coupled_heat_equation}
    \rho c \dot T=k(d)\nabla^2T+T\left(\dfrac{\partial \boldsymbol \sigma}{\partial T}:\dot{\boldsymbol\varepsilon}+\dfrac{\partial K}{\partial T } \dot d+\dfrac{\partial H}{\partial T } \nabla \dot d\right)+Q.
\end{equation}

Eq.~\ref{eq:coupled_heat_equation} can degenerate to the conventional heat conduction equation for heat conduction with an internal heat source. At this point, we adopt the formulation for the quasi-static crack propagation where the transient coupling terms $\dot \varepsilon$ and $\dot d$ vanish. Thereby, in the current implementation, the heat equation takes the form as
\begin{equation}\label{eq:non_coupled_model}
\rho c \dot T=k(d)\nabla^2T+Q. \\
\end{equation}
\subsection{Summary of governing equations}
With the energy terms of the multiphysics problem being defined, the strong form of the quasi-static fracture problem, following the balance laws and constitutive relations described above, can be summarized as follows
\begin{subequations} 
\begin{align}
\text{Momentum balance:}\quad &\nabla \cdot \boldsymbol\sigma=0\\ 
\text{Phase-field equation:}\quad &\dfrac{G_{c}}{c_0\ell_{0}} \omega'(d)-\dfrac{2 G_{c}\ell_{0}}{c_0} \nabla^{2} d- Y=0 \\
\text{Heat equation:}\quad &\rho c \dot T=k(d)\nabla^2T+Q
\end{align} 
\label{eq:strong form govering equations}
\end{subequations}
with the Dirichlet boundary conditions and the Neumann boundary conditions
\begin{equation}
    \left\{\begin{array}{ll}
\boldsymbol{u}(\boldsymbol{x},t)=\boldsymbol{u}^{\ast}(\boldsymbol{x}, t) & \boldsymbol{x} \in \partial\Omega_{\boldsymbol u}, \\
d(\boldsymbol{x}, t) =d^{\ast}(\boldsymbol{x}, t) & \boldsymbol{x} \in \Gamma, \\
T(\boldsymbol{x}, t)={T}^{\ast}(\boldsymbol{x}, t) & \boldsymbol{x} \in \partial\Omega_{T}. \\
\end{array}\right.
\end{equation}
\begin{equation}
    \left\{\begin{array}{ll}
\boldsymbol \sigma\cdot \boldsymbol{n}=\boldsymbol{t}^{\ast} & \boldsymbol{x} \in \partial \Omega_{t}, \\
\boldsymbol{q}\cdot \boldsymbol{n}=\boldsymbol{q}^{\ast} & \boldsymbol{x} \in \partial \Omega_{q}.
\end{array}\right.
\end{equation}
Here, the boundary is partitioned into Dirichlet and Neumann type conditions. Specifically, the boundary is split as follows
\begin{equation}
\partial \Omega=\partial \Omega_{\boldsymbol u} \cup \partial \Omega_{\boldsymbol t}, \quad \partial \Omega_{\boldsymbol u} \cap \partial \Omega_{\boldsymbol t}=\emptyset, \quad \partial \Omega=\partial \Omega_{T} \cup \partial \Omega_{\boldsymbol q}, \quad \partial \Omega_{T} \cap \partial \Omega_{\boldsymbol q}=\emptyset,
\end{equation} 
where $\partial \Omega^ {\boldsymbol u}$, $\partial \Omega^{\boldsymbol t}$,  $\partial \Omega^T$ and $\partial \Omega^{\boldsymbol q}$ is the part of boundary on which the prescribed displacement $\boldsymbol u^{\ast} $, traction $\boldsymbol t^{\ast} $, temperature $T^{\ast}$ and heat flux $\boldsymbol q^{\ast}$ are imposed, respectively, Lastly, the governing equations are supplemented with the following initial conditions. The initial state of the system is considered to be undeformed, undamaged, and unstressed with temperature ${T}_0(\boldsymbol x)$. 
\begin{equation}
    \left\{\begin{array}{ll}
\boldsymbol{u}(\boldsymbol{x}, 0)=\mathbf0 & \boldsymbol{x} \in \Omega, \\
d(\boldsymbol{x}, 0) =0 & \boldsymbol{x} \in \Omega, \\
T(\boldsymbol{x}, 0)={T}_0(\boldsymbol{x}) & \boldsymbol{x} \in \Omega. \\
\end{array}\right.
\end{equation}

\section{Numerical implementation}
\subsection{Finite element discretization}
This section presents the finite element implementation of the model. The weak form is constructed by multiplying the equations in Eq. \ref{eq:strong form govering equations}(a-c) by a corresponding arbitrary test function and integrating them over the domain of the problem. After partial integration, the weak forms take the form:
\begin{subequations} 
\begin{align}
&\int_{\Omega}\boldsymbol \sigma: \nabla {\delta}_{u}dV-\int_{\Omega_t}\boldsymbol{t}^{\ast} {\delta}_{u}dS=0, \\ 
& \int_{\Omega}\dfrac{G_{c}}{\ell_{0}c_0}\omega'(d) \delta_{d}dV+\int_{\Omega}\dfrac{2G_{c} \ell_{0}}{c_0}\nabla d \nabla \delta_{d}dV+\int_{\Omega}g'(d) \mathcal{H} \delta_{d}dV=0, \\
&\int_{\Omega}k\nabla T \nabla \delta_{T}dV -\int_{\Omega_t}k\nabla T \delta_{T}dS +\int_{\Omega}\rho {c}\dot{T}\delta_{T}dV-\int_{\Omega}Q\delta_{T}dV=0.
\end{align} 
\label{eq:weak form govering equations}
\end{subequations}

Utilizing the standard finite element method, the displacement field $\boldsymbol u$, the phase-field $d$ and the temperature field $T$, as well as their first spatial derivatives, are approximated as
\begin{equation}
    \left\{\begin{array}{ll}
\boldsymbol u=\sum N_u^i u_i = \boldsymbol N_u \boldsymbol{u}_e,\quad \boldsymbol\varepsilon=\sum B_u^i u_i = \boldsymbol {B}_u \boldsymbol u_e, \\ 
d=\sum N_d^i d_i =\boldsymbol {N}_{d} \boldsymbol d_e, \quad \nabla d=\sum B_d^i d_i =\boldsymbol B_d\boldsymbol d_e,\\
T= \sum N_T^i T_i =\boldsymbol{N}_{T} \boldsymbol T_e, \quad \nabla T =\sum B_T^i T_i =\boldsymbol{B}_{T}\boldsymbol T_e.\\ 
\end{array}\right.
\end{equation}
Here, $u_i$, $d_i$ and $T_i$ are the nodal values of the displacement, damage and temperature field of node $i$ of element $e$, respectively. $ N_u$, $N_d$, $ N_T$ and $ B_u$, $ B_d$, $ B_T$ denote the shape functions and their derivatives for the displacement, damage field and temperature, respectively, $\boldsymbol N_u$, $\boldsymbol N_d$, $\boldsymbol N_T$ and $\boldsymbol B_u$, $\boldsymbol B_d$, $\boldsymbol B_T$ are the corresponding shape function matrix and derivatives. For a quadrilateral 2D element they are written as
\begin{equation}
    \boldsymbol N_u=\left[
\begin{matrix}
N_1  & 0  &\cdots & N_4  & 0\\
0    & N_1&\cdots & 0    & N_4
\end{matrix}
\right],\quad 
\boldsymbol B_u=\left[
\begin{matrix}
N_{1,x} & 0       &\cdots & N_{4,x}  & 0\\
0       & N_{1,y} &\cdots & 0        & N_{4,y}\\
N_{1,y} & N_{1,x} &\cdots & N_{4,y}  & N_{4,x}
\end{matrix}
\right],
\end{equation}

\begin{equation}
    \boldsymbol N_d=\left[
\begin{matrix}
N_1    &\cdots & N_4 \\
\end{matrix}
\right],\quad 
\boldsymbol B_d=\left[
\begin{matrix}
N_{1,x} &\cdots & N_{4,x}  \\
N_{1,y} &\cdots & N_{4,y}\\
\end{matrix}
\right],
\end{equation}

\begin{equation}
    \boldsymbol N_T=\left[
\begin{matrix}
N_1    &\cdots & N_4 \\
\end{matrix}
\right],\quad 
\boldsymbol B_T=\left[
\begin{matrix}
N_{1,x} &\cdots & N_{4,x}  \\
N_{1,y} &\cdots & N_{4,y}\\
\end{matrix}
\right].
\end{equation}
With the above finite element discretization, we obtain the following equations for the residuals of different fields:
\begin{subequations}\label{eq:residuals}
\begin{align}
\boldsymbol r_{u}=&\int_{\Omega}[\boldsymbol B_{u}]^T\boldsymbol \sigma dV-\int_{\Omega_t} [\boldsymbol N_{u}]^T\boldsymbol{t}^{\ast}dS, \\ 
\boldsymbol r_{d}=&\int_{\Omega}\dfrac{G_{c}}{ \ell_{0}c_0}\omega'(d)[{\boldsymbol N}_{d}]^T \boldsymbol N_{d}dV+\int_{\Omega} \dfrac{2G_c\ell_{0}}{c_0}{[\boldsymbol B_d]}^T\boldsymbol B_{d}dV+\int_{\Omega}g'(d) [\boldsymbol  N_{d}]^T\mathcal{H}dV, \\
\boldsymbol r_{T}=&\int_{\Omega} [\boldsymbol  B_{T}]^T k\boldsymbol  B_{T} dV-\int_{\Omega_t}[\boldsymbol  B_{T}]^T k\boldsymbol  N_{T}dS +\int_{\Omega}\rho c \dot T [\boldsymbol  N_{T}]dV-\int_{\Omega} [\boldsymbol  N_{T}]^TQdV. 
\end{align}
\end{subequations}
\subsection{Staggered solution scheme}
In general, the energy functional of the thermal fracture problem is non-convex with respect to its variables when all the field variables are considered simultaneously. Therefore it is challenging to solve all the unknown variables at the same time utilizing the conventional Newton-Raphson method. However, the problem is convex with respect to the variables $\boldsymbol u$ and $d$ separately when the other is fixed. The latter approach which is also known as the staggered minimization algorithm improves the convergence of the numerical solver.
In this work, the thermo-mechanical coupled problem is first solved in a monolithic way with a fixed crack field. Then the phase-field crack problem is solved with the updated displacement and temperature values. For the $(i+1)$th time step, first, we solve for the nodal displacements and temperature field from the coupled thermo-mechanical problem. At this point, the crack is fixed at $d^{(k)}$ obtained in the previous iteration. Therefore, we first solve for 
\begin{equation}
\begin{aligned}
    \left\{\begin{array}{ll}
\boldsymbol r_{u}^{(i+1)}=&\int_{\Omega}[\boldsymbol B_{u}]^T\boldsymbol \sigma dV-\int_{\Omega_t} [\boldsymbol N_{u}]^T\boldsymbol{t}^{\ast}dS, \\ 
\boldsymbol r_{T}^{(i+1)}=&\int_{\Omega} [\boldsymbol  B_{T}]^T k\boldsymbol  B_{T} dV-\int_{\Omega_t}[\boldsymbol  B_{T}]^T k\boldsymbol  N_{T}dS+\int_{\Omega}\rho c \dot T^{(i+1)} [\boldsymbol  N_{T}]dV-\int_{\Omega} [\boldsymbol  N_{T}]^TQdV, \\ 
\end{array}\right.
\end{aligned}
\end{equation}
where $\boldsymbol \sigma=\boldsymbol\sigma(\boldsymbol u^{(i+1)} , d^{(i)},T^{(i+1)})$, $k=k(d^{(i)})$ and $G_c=G_c(T^{(i+1)})$. 
This standard thermo-mechanical problem can be solved by the Newton method, with the equation 
\begin{equation}
    \left[\begin{array}{cc}
\boldsymbol{K}_{ u  u} & \boldsymbol{K}_{ u T} \\ 
\boldsymbol{K}_{ T  u} & \boldsymbol{K}_{T T}
\end{array}\right]
\left[\begin{array}{l}
\Delta \boldsymbol u \\ 
\Delta T
\end{array}\right]=
\left[\begin{array}{l}
{\boldsymbol r}_{u}^{(i+1)} \\ 
{\boldsymbol r}_{T}^{(i+1)}
\end{array}\right].
\end{equation}
Next, we solve the nodal unknowns for the crack problem with the updated nodal displacement and nodal temperature $\boldsymbol u^{(i+1)},T^{(i+1)}$ i.e,
\begin{equation}
    \boldsymbol r_{d}^{(i+1)}=\int_{\Omega}\dfrac{G_{c}}{ \ell_{0}c_0}\omega'(d)[{\boldsymbol N}_{d}]^T \boldsymbol N_{d}dV+\int_{\Omega} \dfrac{2G_c\ell_{0}}{c_0}{[\boldsymbol B_d]}^T\boldsymbol B_{d}dV+\int_{\Omega}g'(d) [\boldsymbol  N_{d}]^T\mathcal{H}^{(i+1)}dV.
\end{equation}
The linearization of the above relation yields
\begin{equation}
    \boldsymbol K_{d d}\Delta d=\boldsymbol r_d^{(i+1)}
\end{equation}

Based on the above explanations, the algorithm for thermal fracture is summarized in Algorithm 1.

\begin{center}
\setlength\fboxsep{0pt}
\vskip-\lastskip%
{%
\begin{minipage}{\linewidth}
\vspace{-12pt}
      \begin{algorithm}[H]
        \caption{Staggered minimization algorithm at time interval [$t_n$, $t_{n+1}$]}
        \label{alg:alg10-a} 
        \begin{algorithmic}[1] 
          \State Inputs: solutions of temperature field $T_{n}$, displacement field $\boldsymbol u_{n}$, crack field $d_{n}$ at former time step $t_n$.
          \State Outputs: temperature field $T_{n+1}$, displacement field $\boldsymbol u_{n+1}$, crack field $d_{n+1}$ at current time step $t_{n+1}$.
          \State Set $i=1$, Tolerance $=$ 1e-8. 
          \State Set $T_{n+1}^{(0)}$  $\leftarrow$ $T_{n}$, $u_{n+1}^{(0)}$  $\leftarrow$ $u_{n}$, $d_{n+1}^{(0)}$  $\leftarrow$ $d_{n}$ 
          \Repeat
          \State Compute $T_{n+1}^{(i)}$, $\boldsymbol u_{n+1}^{(i)}$ and fix $d_{n+1}^{(i-1)}$
          
          \State Compute $d_{n+1}^{(i)}$ and fix $T_{n+1}^{(i)}$, $\boldsymbol u_{n+1}^{(i)}$ 
          \State Check the irreversibility constraint $d_{n+1}^{(i)} \geq d_{n}$
          \State i $\leftarrow$ i+1
          \Until $\mid{d_{n+1}^{(i)}-d_{n+1}^{(i-1)}}\mid$ $\leq$ Tolerance
          \State Update solutions $T_{n+1} \leftarrow T_{n+1}^{(i)}$,~$u_{n+1} \leftarrow u_{n+1}^{(i)}$, and $d_{n+1} \leftarrow d_{n+1}^{(i)}$
        \end{algorithmic}
      \end{algorithm}
      \end{minipage}%
      }
\end{center}

The model is numerically implemented by the Finite Element Method within the framework of Multiphysics Object-Oriented Simulation Environment (MOOSE)~\cite{permann2020moose}. It is worth noting that the Automatic Differentiation (AD) capabilities in MOOSE are utilized here, which is a symbolic differentiation method \cite{Lindsay2021}. It applies the chain rule and propagates derivatives to elementary operations at every step. AD offers a very accurate Jacobian at a relatively small overhead cost. Thus there is no need to compute Jacobian by hand which is arduous and prone to errors in the context of multiphysics problems, shifting the burden of computing the derivatives of the complex known expressions for the free energies from the user to the software. MOOSE employs the DualNumber class from the MetaPhysicL package to enable forward-mode AD capabilities \cite{malaya2013masa}.

\section{Benchmark examples}
Before the developed model is applied for hot cracking simulations in the complex situation during AM in the next section, simulation results on a few benchmark thermal fracture problems are presented in this section to demonstrate and check the reliability of the proposed model. Thereby the single edge notched tension test with the thermo-mechanical loading and the quenching test of a solid plate are studied, respectively.  

\subsection{Single edge notched tension test}
\begin{figure}[H]
    \centering
    \includegraphics[width=\textwidth]{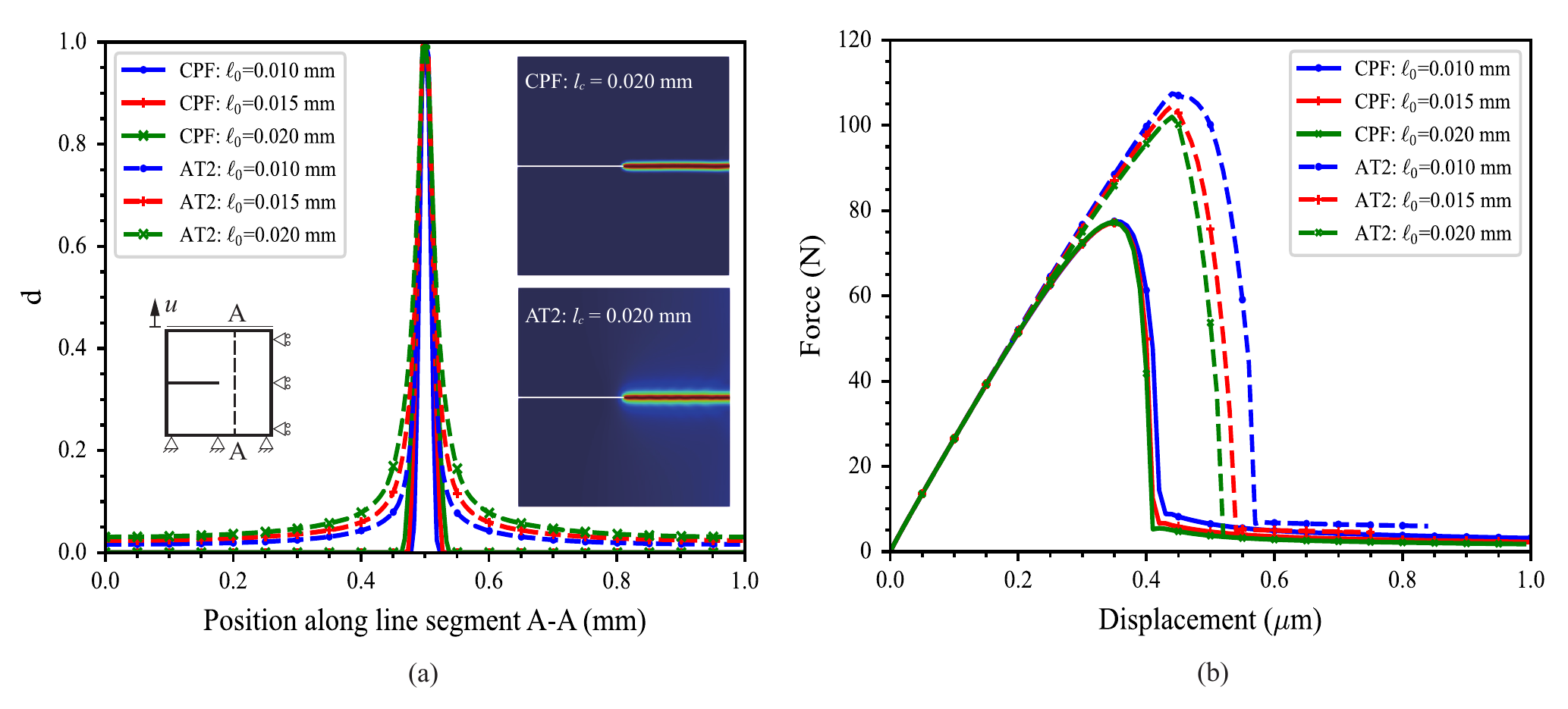}
    \caption{Results of single edge notched tension test under thermo-mechanical boundary conditions. (a) Comparison of damage level of AT2 model and CPF model with different $\ell_{0}$. (b) Reaction force-displacement curve with different $\ell_{0}$.}
    \label{fig:d_profile_wu_power}
\end{figure}

We start by investigating the single edge notched tension test, which has become canonical in the phase-field models for fracture. Consider a square plate of 1mm $\times$ 1mm with an initially horizontal edge crack extending to the middle of the specimen. The geometry and boundary conditions are shown in Fig. \ref{fig:d_profile_wu_power}(a). The bottom edge is fixed while a vertical displacement is applied to the top edge. To include the temperature field in the thermal fracture model, the initial temperature of the plate is set as 300 K. The outer boundaries are treated as adiabatic.  The domain is discretized into the unstructured mesh, and the mesh is refined along the anticipated crack path to assure the fracture phase-field variable in the localized band is well resolved. This point is essential to accurately capture the evolution of the crack field in the simulations.

\begin{table}[H]
      \centering
      \caption{Material parameters for single edge notched tension test}
        \begin{tabular}{c c c c c c c c c} 
        \hline
        $E$ (GPa) & $\nu$ & $G_c$ (J/m$^2$) & $\sigma_u$ (MPa) & $\rho$ (kg/m$^3$) & $k$ (W/mK)  & $c_p$ (J/kgK)  & $\alpha$ \\        \hline
        340      & 0.22   & 42.47          & 180        & 2450          & 300          & 0.775        &  8.0$\times$ $10^{-6}$ \\        \hline
        \end{tabular}
        \label{Table:Material parameters for single edge notched tension test}
\end{table}

To compare the difference between AT2 model and CPF model and the $\ell_{0}$ insensitivity of the latter model, three different length scale parameters $\ell_{0}$ = 0.010 mm, 0.015 mm, and 0.020 mm are used, both for the AT2 model and the CPF model. The  refined mesh size is set to be 0.005 mm. Thus the ratio of length scale to mesh size is 2, 3, and 4 respectively. The fixed time step is 0.010 ms and the total simulation time is 1.0 ms. The material properties used for these examples are listed in Table. \ref{Table:Material parameters for single edge notched tension test}.

The AT2 model and CPF model are compared here to check the convergence of the results concerning the length scale parameter $\ell_{0}$. In Fig. \ref{fig:d_profile_wu_power}(a), the crack patterns calculated using the CPF model and AT2 model with $\ell_{0}$ = 0.020 mm are shown, respectively. As $\ell_{0}$ controls the width of the diffusive zone, with the increase of $\ell_{0}$ the damaged zone becomes wider. Fig. \ref{fig:d_profile_wu_power}(a) also shows that the damaged zones of the CPF model are more compact compared with the AT2 model, as the half bandwidth of the former model is $\pi \ell_{0}$/2 while for the latter is infinity \cite{wu2018length, hu2020phase}. This result also verifies there is a threshold of damage initiation for the CPF model.
The corresponding load versus displacement curves are plotted in Fig. \ref{fig:d_profile_wu_power}(b). The results show that the global responses of the CPF model are almost independent of $\ell_{0}$ while for the AT2 model the peak load decreases with the increase of the length scale. The former observation confirms the insensitivity of the results from the CPF model concerning the length scale parameter under thermo-mechanical conditions.

\begin{figure}[H]
\centering
\includegraphics[width=1.0\textwidth]{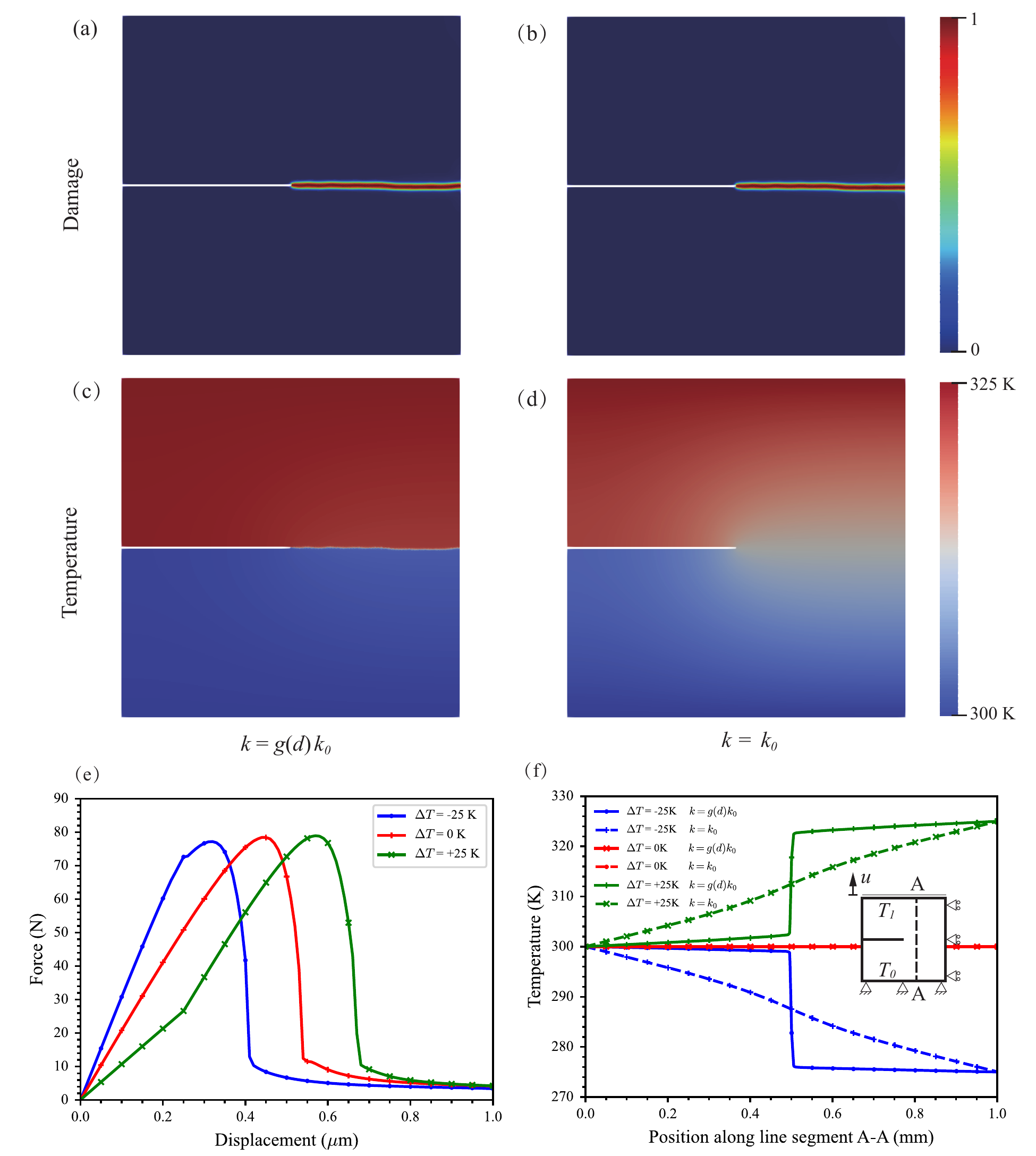}
\caption{
Single edge notched tension test under thermo-mechanical loading. Crack patterns (a and b) and temperature profiles (c and d)  for case 3 with and without thermal conductivity degradation. (e) Comparison of the reaction force-displacement curves for different thermal loading and (f) Temperature profiles along A-A for different thermal loading with and without thermal conductivity degradation.}  
\label{fig:crack_T_changeT}
\end{figure}

We further study the effect of thermal loading on the crack patterns and temperature distributions. Three cases are considered here. In case 1, we keep the temperature of the bottom and the top edges at 300 K. In case 2, we cool down the top edge from the beginning to 0.25 ms at a ratio of 1.0e5 K/s to 275 K and then keep the temperature constant. Finally, in Case 3, we heat up the top edge with the same speed to 325 K and then keep it constant. Other settings of the simulations are the same as before. The comparisons of these results are shown in Fig. \ref{fig:crack_T_changeT}. It is observed that compared with case 1 where the constant thermal loading is included and the crack propagates horizontally to the right edge of the specimen, the crack develops slightly downwards when the temperature is increasing on the top edge, while upwards when the temperature is decreasing. Take case 3 for illustration here. Fig. \ref{fig:crack_T_changeT}(a) and (b) shows the crack patterns of case 3 with degraded and constant thermal conductivity. The reaction force-displacement curves are plotted in Fig. \ref{fig:crack_T_changeT}(e). From the curves, it is apparent the onset of the crack is delayed when the top edge is heated up compared with the fixed temperature while the crack propagates earlier when cooling down the top edge. When the temperature goes up, the specimen expands and the thermal strain is positive, which leads to a smaller elastic strain with fixed displacement and a smaller driving force for crack propagation compared with a uniform temperature field. 
Meanwhile, the fracture energy $G_c$ is a function of temperature and it decreases when the temperature increases at the temperature interval of interest  (see Eq. \ref{Eq:Gc(T)}). According to Griffith's theory \cite{griffith1921vi}, the fracture strength of material for plane strain problem is defined as
\begin{equation}
    \sigma_f=\sqrt{\dfrac{2E\gamma_s}{\pi l (1-\nu^2)}}=\sqrt{\dfrac{EG_c}{\pi l(1-\nu^2)}}
    \label{Eq:fracture_strength}
\end{equation}
where $\gamma_s$ is the fracture surface energy density. Therefore, the decreasing $G_c$ also leads to the decreasing $\sigma_f$ for case 3. However, the value changes very slightly and cannot compensate for the influence of thermal strain. Therefore, a larger displacement is needed for the onset of crack initiation. While for case 2 with decreasing temperature, the contraction of the specimen results in negative thermal strain. Correspondingly, the elastic strain and the driving force are greater. Therefore the onset of the crack only needs a smaller displacement, despite that $G_c$ and $\sigma_f$ increases slightly.

Note that the thermal conductivity is degraded as the crack develops. This effect is considered to avoid non-physical heat transfer happening in the fully-cracked region. The comparisons of the temperature profile of case 3 with degraded and constant thermal conductivity are depicted in \ref{fig:crack_T_changeT}(c) and (d). It shows that when there is no thermal conductivity degradation with the phase=field crack, the temperature field changes smoothly even in the cracked regions. However, as the thermal conductivity is degraded with crack, the temperature is not continuous across the cracked regions. The phenomenon can also be observed in Fig. \ref{fig:crack_T_changeT}(f), the temperature profiles along the line segment A-A at the final time step for the three cases are depicted. Interestingly enough, when the thermal conductivity is degraded with the phase=field crack, a sharp temperature jump is observed for case 2 and case 3. For case 1, because the temperature at the top edge and bottom edge are both fixed at 300K, the temperature does not change. In contrast, the smooth temperature field change is observed when no thermal conductivity degradation is considered.

\begin{figure}[H]
    \centering
    \includegraphics[width=1.0\textwidth]{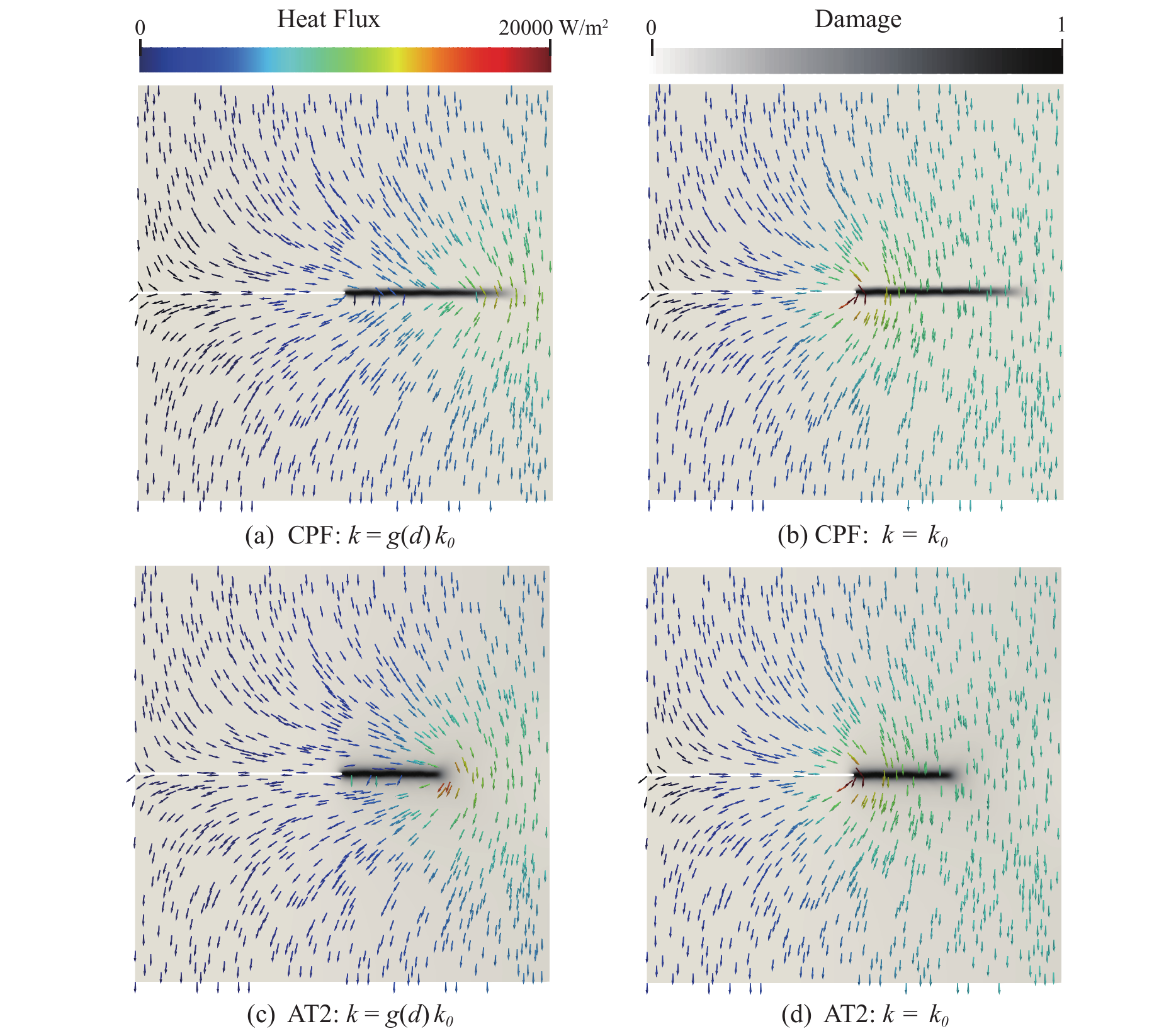}
    \caption{Snapshots of heat flux for cases with degraded (left column) and constant (right column) thermal conductivity for CPF model (a and b) and AT2 model (c and d).}  
    \label{fig:flux}
\end{figure}

To further illustrate the influence of the degraded thermal conductivity, Fig. \ref{fig:flux} depicts the heat flux magnitudes and directions for the  case 3 with degraded and constant thermal conductivity. The initial heat flux directions are identical for the two scenarios. Because of the temperature gradient between the top edge and the bottom edge, heat flux goes from high-temperature regions to lower-temperature regions. In the right half of the specimen containing no crack, the heat flux goes almost vertically downwards and perpendicular to the pre-crack direction, while for the other half the heat flux tries to bypass the notch tip. However, with different thermal conductivity being considered, the heat flux becomes different when the crack propagates forward, as shown in Fig. \ref{fig:flux}(a) and (b). It is observed that when thermal conductivity is degraded with crack (see Eq. \ref{eq:thermal_condutivity}), the heat flux directions change at the crack tip while it is still perpendicular to the crack for the constant thermal conductivity case. Because of the cohesive nature of the CPF model, the damage level at the crack tip is more diffusive. And also the residual conductivity still exists for the parameter $\xi$, so the thermal conductivity is not fully degraded. Thus there is still some residual heat flux crossing the crack tip. This degradation phenomenon is also observed when the AT2 model is used, see Fig. \ref{fig:flux}(c) and (d). When the degraded thermal conductivity is considered, its value approaches zero as soon as the crack is fully developed, meaning the material is not continuous physically. Thus there is no heat transfer happening in the fully cracked regions, and no heat flux across these regions either. Instead, the heat flux arrows circumvent the crack tip (Fig. \ref{fig:flux}(a) and (c)). On the contrary, when the thermal conductivity is not degraded with crack, i.e. constant crack, the heat flux directions don't change at all and the flux still crosses the cracked regions (Fig. \ref{fig:flux}(b) and (d)), which is not physically correct. It is also observed that the magnitudes of the heat flux increase at the crack tip, since the flux concentrates at this region. Note that one can also consider other sources of heat transfer (e.g. convection or even radiation) at the damaged zone or where we have the discontinuity in the displacement field. In the current work, we are first restricted to heat transfer through conduction in solids. 

\subsection{Quenching test}
In this section, the quenching tests are presented to further verify the phase-field fracture model. In the experimental side of this test, a ceramic plate with an initially high temperature is subjected to a cool water bath, and a series of parallel cracks are formed in the ceramics \cite{wang2020phase, jiang2012study}. 

In the numerical example, a ceramic slab of 50 mm $\times$ 10 mm with a high initial temperature $T_0$ is considered. The ambient temperature $T_a$ is lower, so there is a temperature difference $\Delta$T. For computational efficiency, only a quarter of the whole rectangle plate is modeled here. The symmetry boundary conditions are therefore applied on the corresponding edges, where the adiabatic condition is applied (i.e. no flux goes through these edges). The horizontal displacement is fixed on the right edge while the vertical displacement on the top edge is fixed, respectively. The remaining edges are subjected to quenching through heat conduction. The material properties are taken from \cite{mandal2021fracture} and are listed in the Table. \ref{Table:Material parameters used for quenching test}. In this quenching test, $T_0$ = 300 $^{\circ}$C, $T_a$ = 20 $^{\circ}$C and thus $\Delta$T = 280 $^{\circ}$C. The length scale here $\ell_{0}$ = 0.10 mm with the fine mesh size $h=\ell_{0}/4$. The fixed time step is $\Delta t$ = 0.10 ms and the total simulation time is 200 ms. 

 \begin{table}[H]
      \centering
      \caption{Material parameters used for the quenching test.}
        \begin{tabular}{c c c c c c c c c} 
        \hline
        $E$ (GPa) & $\nu$ & $G_{c}$ (J/m$^2$) & $\sigma_u$ (MPa) & $\rho$ (kg/m$^3$) & $k$ (W/mK)  & $c_p$ (J/kgK)  & $\alpha$ \\        \hline
        370      & 0.3   & 42.47          & 180        & 3980          & 31          & 880        &  $7.5\times 10^{-6}$ \\        \hline
        \end{tabular}
        \label{Table:Material parameters used for quenching test}
\end{table}

The temperature field, as well as crack development of the quenching test are simulated. At the beginning of the quenching test, the outer boundaries of ceramics are damaged due to the high tensile thermal stress induced by the temperature variation. Under the influence of the thermal shock, cracks initiate and propagate almost uniformly with an equal spacing, which are almost perpendicular to the boundaries. Initially, the cracks propagate quite rapidly, as they propagate to the inner of the specimen. Then the propagation speed decreases gradually with the release of the thermal stress. Some cracks get arrested at a short length because the declining strain energy is unable to support all the cracks to further propagate simultaneously. Therefore, the remaining cracks gain more driving force to keep propagating further. The process repeats once again until the final crack pattern forms. For a more detailed description of the experimental quenching process, readers are referred to \cite{jiang2012study}.

\begin{figure}[H]
    \centering
    \includegraphics[width=1.0\textwidth]{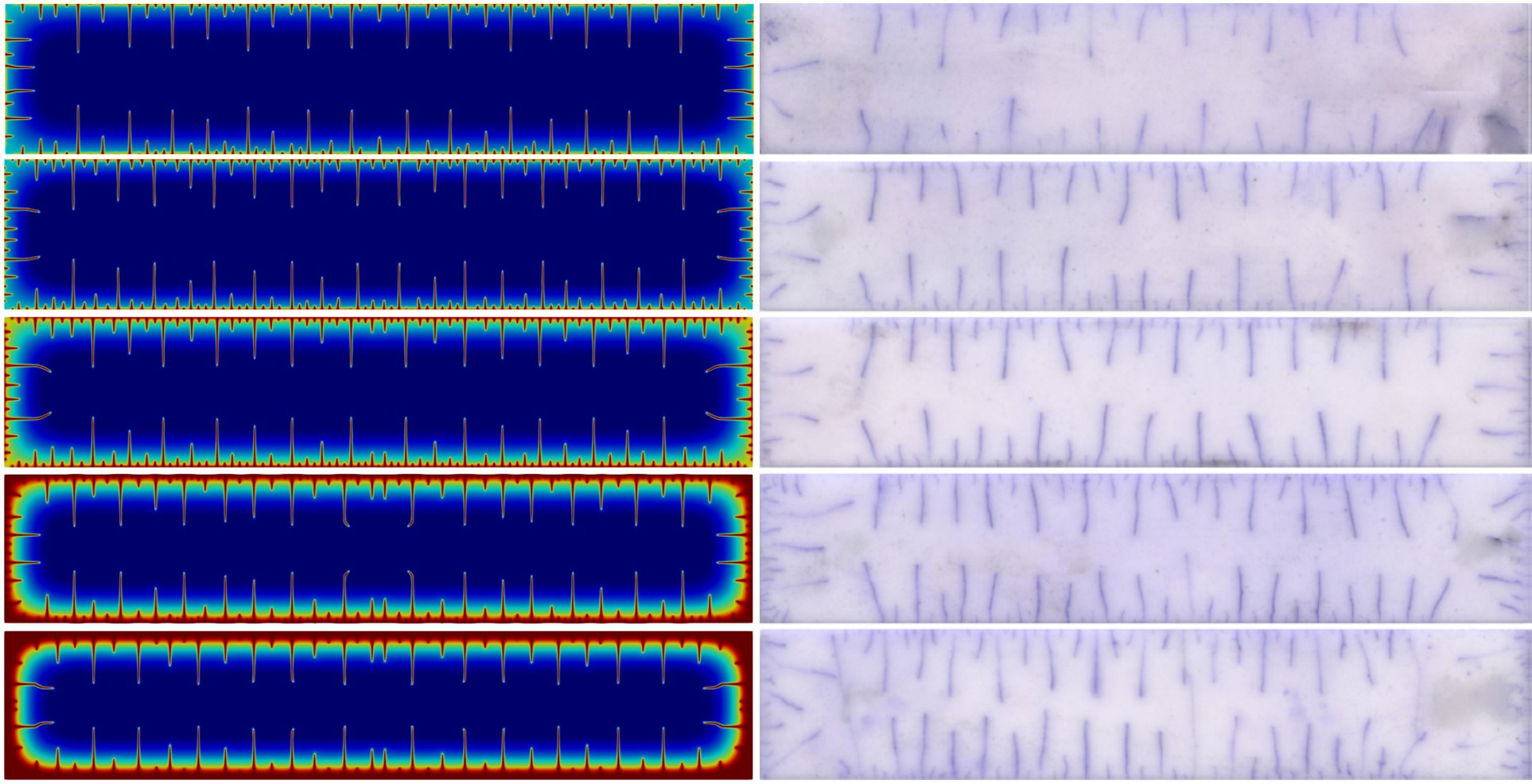}
    \caption{Quenching tests with different initial temperatures $T_0$. Comparison of crack patterns between numerical and experimental results. Reprinted from \cite{jiang2012study} with permission. }
    \label{fig:crack pattern with different initial temperature}
\end{figure}

The comparison between the numerical result and experimental result of the quenching test for the whole specimen with an initial temperature of 300 $^{\circ}$C is shown in the first row in Fig. \ref{fig:crack pattern with different initial temperature}. The crack pattern for the whole ceramic specimen is obtained by using symmetry conditions at the top edge and right edge of the quarter plate in post-possessing. From the comparison, it is apparent that the results obtained from the proposed thermal fracture model shows good agreement with the experiment results.

To further study the influence of the initial temperature of the ceramics on the final crack pattern, more quenching tests with different initial temperatures, i.e. 350 $^{\circ}$C, 400 $^{\circ}$C, 500 $^{\circ}$C and 600 $^{\circ}$C are conducted. The material parameters and model settings are the same as before. The comparison of the simulation results with the experimental observations \cite{jiang2012study} with different initial temperatures $T_0$ is shown in Fig. \ref{fig:crack pattern with different initial temperature}.
From the comparison, the thermal shock crack patterns (spacing, length, height hierarchy, and periodicity) are similar for different initial temperatures $T_0$. The crack mechanism like crack initiation and propagation follows the same pattern as described before. However, the crack pattern still keeps evolving with the increase of $T_0$. The boundaries of the ceramics have a higher level of damage with increasing $T_0$. More cracks at the boundaries are formed as a result of higher thermal stress. In addition, the crack spacing also gets smaller and the longer cracks propagate even longer to the central part of the specimen. The experimental observations can also be reflected by the simulation results. 

\subsection{Hot cracking in additive manufacturing}
In this section, we present the first attempt to numerically study the hot cracking phenomenon during the AM process with the proposed phase-field model for thermal fracture. 
As thermal initial condition, we apply a simple analytical solution for the thermal profile around the melt pool and the numerically calculated thermal profile by a phenomenological thermal PBF model, respectively. The latter allows a more high-fidelity temperature field during PBF process, and the parameter study on different process parameters. 


For explaining the simulation setup, we need to introduce also the solidification strain. In PBF, the solid-state powders melt when the temperature exceeds the melting temperature. It then resolidifies when the temperature is lower than the liquidus temperature. Most metals and alloys contract on solidifying, and the liquid-solid contraction leads to the volume change, which adds the phase transformation strain to the total strain calculation, as shown in Fig. \ref{fig:thermal profile}(a). Therefore, the mechanism of the solidification shrinkage is different from that of thermal strain induced by temperature change \cite{kalpakjian1984manufacturing}. Meanwhile, the contribution of solidification shrinkage is relatively significant and has to be considered in the calculations. For example, the solidification shrinkage for aluminum is 6.6\%, equivalent to 2.2\% of the linear contraction, which is about 50\% greater than the thermal contraction of cooling from the melting temperature to room temperature (about 1.5\%) \cite{feng1994computational}. Therefore, the solidification shrinkage would play a significant role in the strain and stress field, particularly in the vicinity of the melting pool. Interestingly enough, this phenomenon is usually overlooked by most of researchers in numerical simulation of AM process.

In this work, the solidification shrinkage strain $\boldsymbol \varepsilon_{SS}$ is simulated utilizing the effective thermal expansion coefficient. For alloys, solidification happens within a certain temperature range, i.e. the liquidus temperature $T_L$ and solidus temperature $T_S$. The shrinkage is proportional to the change of solid fraction. Therefore, it is assumed to be linearly distributed in the temperature range and can be treated as an additional thermal expansion term caused by temperature changes \cite{feng1994computational}. Here we assume the effective thermal expansion coefficient $\tilde{\alpha}(T)$ takes the following form:

\begin{equation}
    \tilde{\alpha}(T)=\begin{cases} 
    \alpha_T, & T \leq T_S \\
    \alpha_{T}+\alpha_{SS},& T_S\leq T \leq T_L \\ 
    \alpha_T, & T_L \leq T\end{cases}, 
\end{equation}
where $\alpha_{SS} = \dfrac{\boldsymbol \varepsilon_{SS}}{T_L-T_S}$. Correspondingly, the thermal strain is obtained via:

\begin{equation}
    \boldsymbol{\varepsilon}_{T}=\tilde{\alpha}(T)(T-T_0)=\begin{cases}
    \alpha_T(T-T_0),& T< T_S\\
    \alpha_T(T-T_0)+\alpha_{SS}(T-T_S), &T_S\leq T < T_L\\
    \alpha_T(T_L-T_0)+\boldsymbol{\varepsilon}_{SS}, &T_L \leq T
\end{cases}
\end{equation}

\subsubsection{Hot cracking simulation by using analytical elliptic temperature profile}



\begin{figure}[H]
    \centering
    \includegraphics[width=1.0\textwidth]{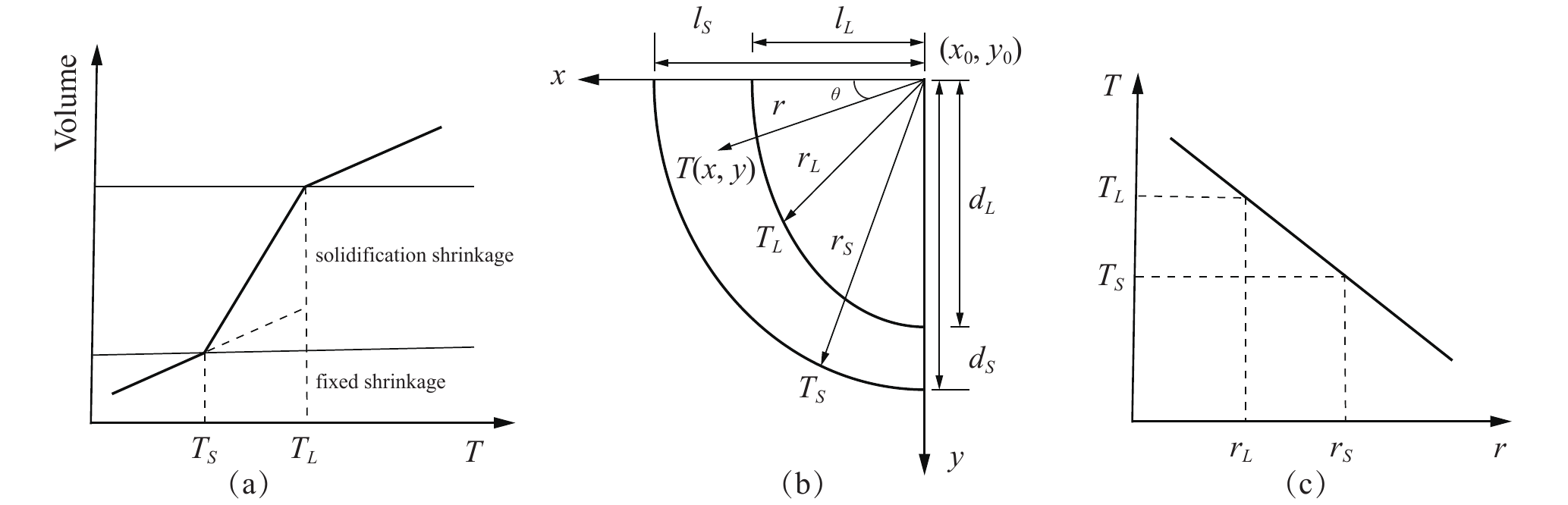}
    \caption{(a) Volume change with temperature \cite{kalpakjian1984manufacturing} and solidification shrinkage between $T_S$ and $T_L$. (b) Schematic of the elliptic temperature field approximation. (c) Linear interpolation of temperature profile.}
    \label{fig:thermal profile}
\end{figure}

We start with the analytical elliptic solution of the temperature profile in the vicinity of the melting pool \cite{elahi2022multiscale}, half of which is shown in Fig.~\ref{fig:thermal profile}(b). For a specific point $(x,y)$ in Cartesian coordinates within the cross section perpendicular to the scanning direction, $x$ and $y$ are the distance from the current location to the center line and the top surface of the melting pool, respectively. The point can also be determined by polar coordinates $(r,\theta)$, which are calculated by
\begin{equation}
r=\sqrt{x^2+y^2}, \quad
\theta=\tan ^{-1}\left|\frac{y-y_{0}}{x-x_{0}}\right|
\end{equation}
where $r$ denotes the distance to the center of the ellipses $(x_0,y_0)$, and $\theta$ is the angle starting from the top surface to the current position, as shown in Fig.~\ref{fig:thermal profile}(b). The temperature is calculated by linear interpolation between liquidus temperature $T_L$ and solidus temperature $T_S$ as

\begin{equation}\label{eq:linear_interpolation}
    T(x,y)=T(r,\theta)=T_L+c (T_S-T_L)\dfrac{r-r_L(\theta)}{r_S(\theta)-r_L(\theta)}.
\end{equation}
Here, $r_{L}(\theta)$ and $r_{S}(\theta)$ indicate the respective isolines of $T=T_L$ and $T=T_S$ ellipses. They are function of $\theta$, and are calculated by 

\begin{equation}
\begin{split}
r_{L}(\theta)=\sqrt{\frac{\left(l_{L} d_{L}\right)^{2}}{\left(d_{L} \cos (\theta)\right)^{2}+\left(l_{L} \sin (\theta)\right)^{2}}}
\end{split}
\end{equation}
\begin{equation}
\begin{split}
r_{S}(\theta)=\sqrt{\frac{\left(l_{S} d_{S}\right)^{2}}{\left(d_{S} \cos (\theta)\right)^{2}+\left(l_{S} \sin (\theta)\right)^{2}}}.
\end{split}
\end{equation}

The length parameters $(l_L, l_S)$ and the depth parameters $(d_L,d_S)$ are corresponding to liquidus and solidus isotherms, respectively. Given that there is a large temperature gradient in the vicinity of the melting pool, a coefficient $c$ is included in Eq.~\ref{eq:linear_interpolation}. The approximated temperature field is schematically shown in Fig. \ref{fig:thermal profile}(c).

\begin{figure}[H]
    \centering
    \includegraphics[width=1.0\textwidth]{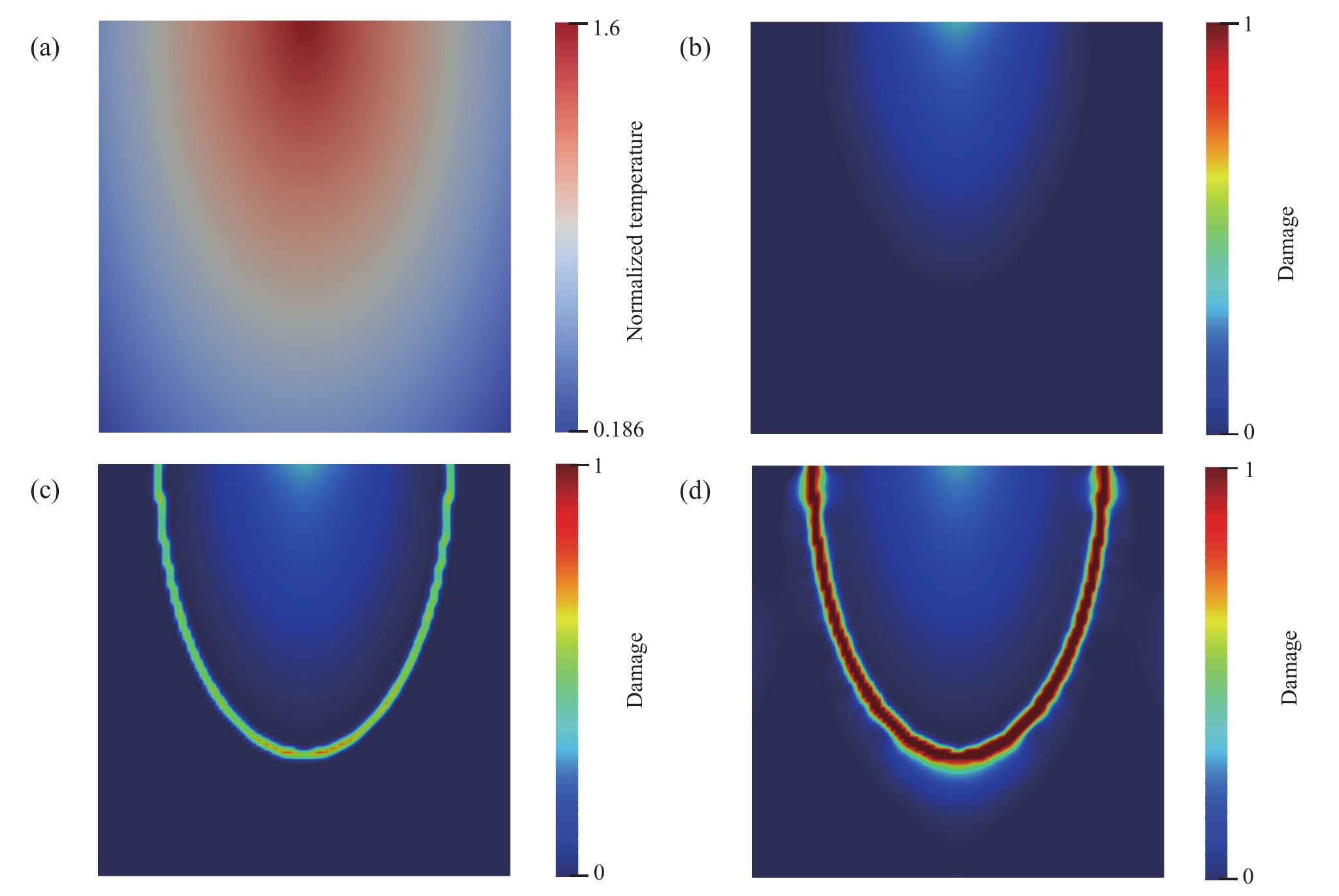}
    \caption{Hot cracking of single track of PBF with aspect ratio to be 1.0. (a) Normalized temperature field obtained by linear interpolation approximation, (b)-(d) Hot cracking patterns with increasing solidification shrinkage strain, $\alpha_{SS}=0, 4\alpha_T$ and $26.7\alpha_T$, respectively.}
    \label{fig:hot_crack_T_d}
\end{figure}

\begin{figure}[H]
    \centering
    \includegraphics[width=1.0\textwidth]{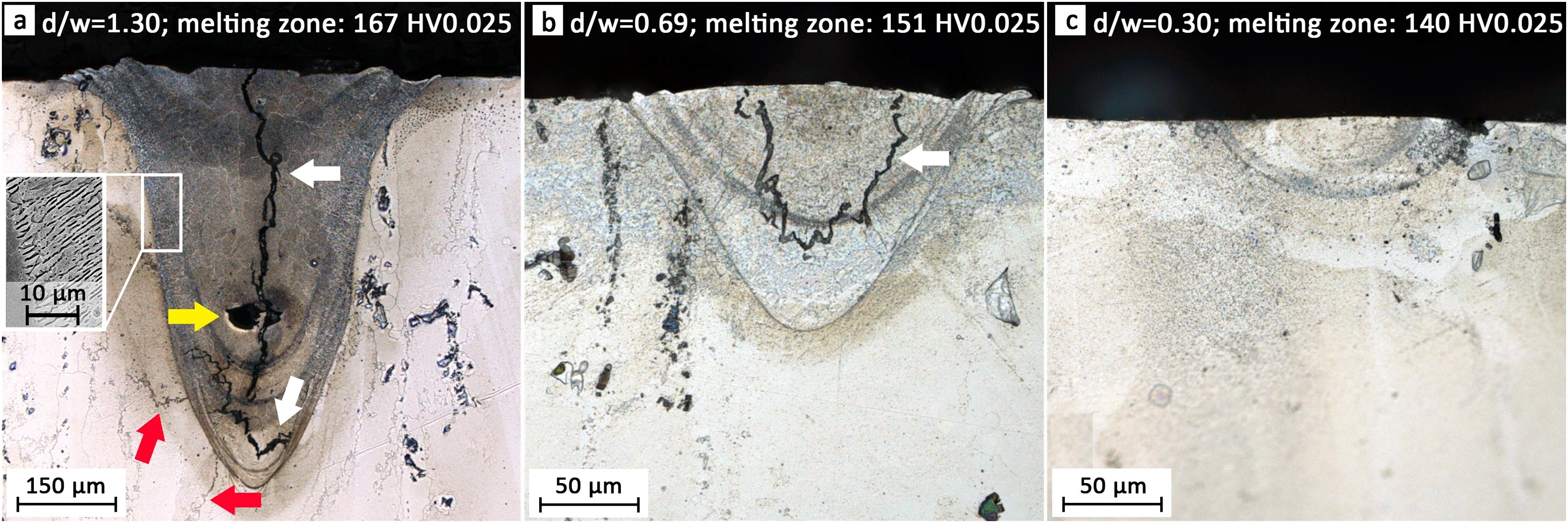}
    \caption{Hot cracking of single track of PBF with different power densities in experimental observations. The image is reprinted from \cite{stopyra2020laser} under the terms of the Creative Commons CC-BY license. }
    \label{fig:hot_crack_experiment}
\end{figure}

In this example, a cross section of 10 mm$\times$10 mm perpendicular to the scanning direction is simulated. Since it is mechanically constraint along the scanning direction, the plane strain assumption is taken. The liquidus and solidus temperatures are assumed to be 890 K and 900 K, respectively, and the reference temperature is 1000 K. The coefficient $c$ is set to 10. The aspect ratio of the melting pool equals to 1.0, with parameters $l_L$ , $l_S$ , $d_L$ and $d_S$ to be 3.5 mm, 4.0 mm, 7.0 mm and 8.0 mm. The normalized temperature field obtained from the analytical solution is shown in Fig.~\ref{fig:hot_crack_T_d}(a). The predicted hot cracking patterns with the increasing solidification shrinkage strain are shown in Fig.~\ref{fig:hot_crack_T_d}(b)-(d). When the solidification shrinkage strain is not considered ($\alpha_{SS} = 0$), a small level of damage is observed only in the region of the melting pool where the temperature is quite high. Outside these regions, there is no damage, as shown in Fig.~\ref{fig:hot_crack_T_d}(b). As the value of solidification shrinkage increases ($\alpha_{SS} = 4\alpha_T$), as shown in Fig.~\ref{fig:hot_crack_T_d}(c), a circumferential crack starts to form which is mainly localized in the regions where the temperature is between the interval of the liquidus and solidus isotherms (see also the temperature field profile). As illustrated above, the material goes through a phase transformation in this temperature range, which causes volume change and relatively larger solidification shrinkage strain compared with thermal strain in other regions. When the solidification shrinkage strain is significant enough ($\alpha_{SS} = 26.7\alpha_T$), the circumferential crack is more obvious, as can be seen in Fig.~\ref{fig:hot_crack_T_d}(d). The latter observation is in agreement with the experimental results observed in the laser PBF process of alloy \cite{stopyra2020laser}. See also the middle part of Fig.~\ref{fig:hot_crack_experiment}.

\begin{figure}[H]
    \centering
    \includegraphics[width=1.0\textwidth]{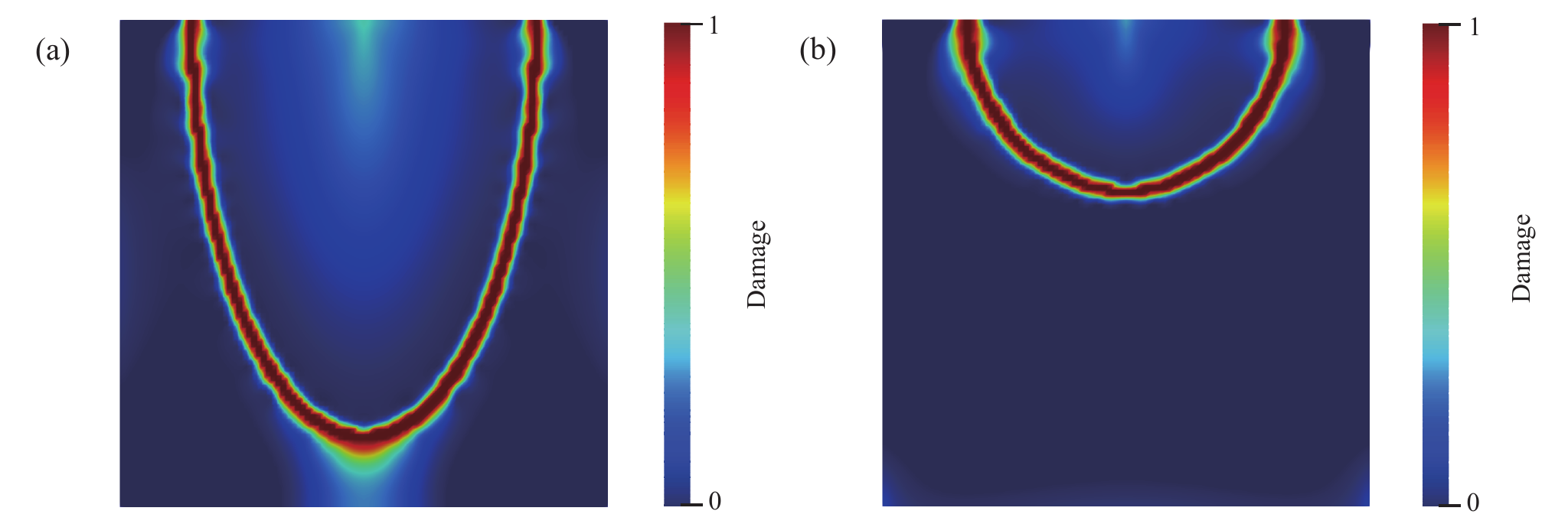}
    \caption{Hot cracking patterns of a single track of PBF with different shapes of the melting pool. (a) Keyhole mode with an aspect ratio of 1.2. (b) Conduction mode with an aspect ratio of 0.5.} 
    \label{fig:hot_crack_meltingpool}
\end{figure}

In order to study the influence of the shape of melting pool resulting from different AM process parameters, two more melting pools with larger (1.2) and smaller (0.5) aspect ratios are studied. For the former one, the melting pool is deeper, which results from the larger energy density and it is often referred to as keyhole mode in AM \cite{tenbrock2020influence}. For the latter one where the energy density is smaller and the melting pool is shallower, it is called conduction mode. The temperature field is obtained by the same method illustrated above. The results are shown in Fig. \ref{fig:hot_crack_meltingpool}. The crack patterns resemble to the result shown in Fig. \ref{fig:hot_crack_T_d}(d), and only circumferential crack shows in the melting pool. Results show the solidification shrinkage is responsible for the circumferential crack of the hot cracking pattern. Hereby the central cracking is not much visible, which can be due to the fact that the inaccuracy of the analytical thermal profile assumed above. As it can be seen in the next subsection, a central crack can be formed as a consequence of high thermal gradient at high energy density.

\begin{figure}[H]
    \centering
    \includegraphics[width=1.0\textwidth]{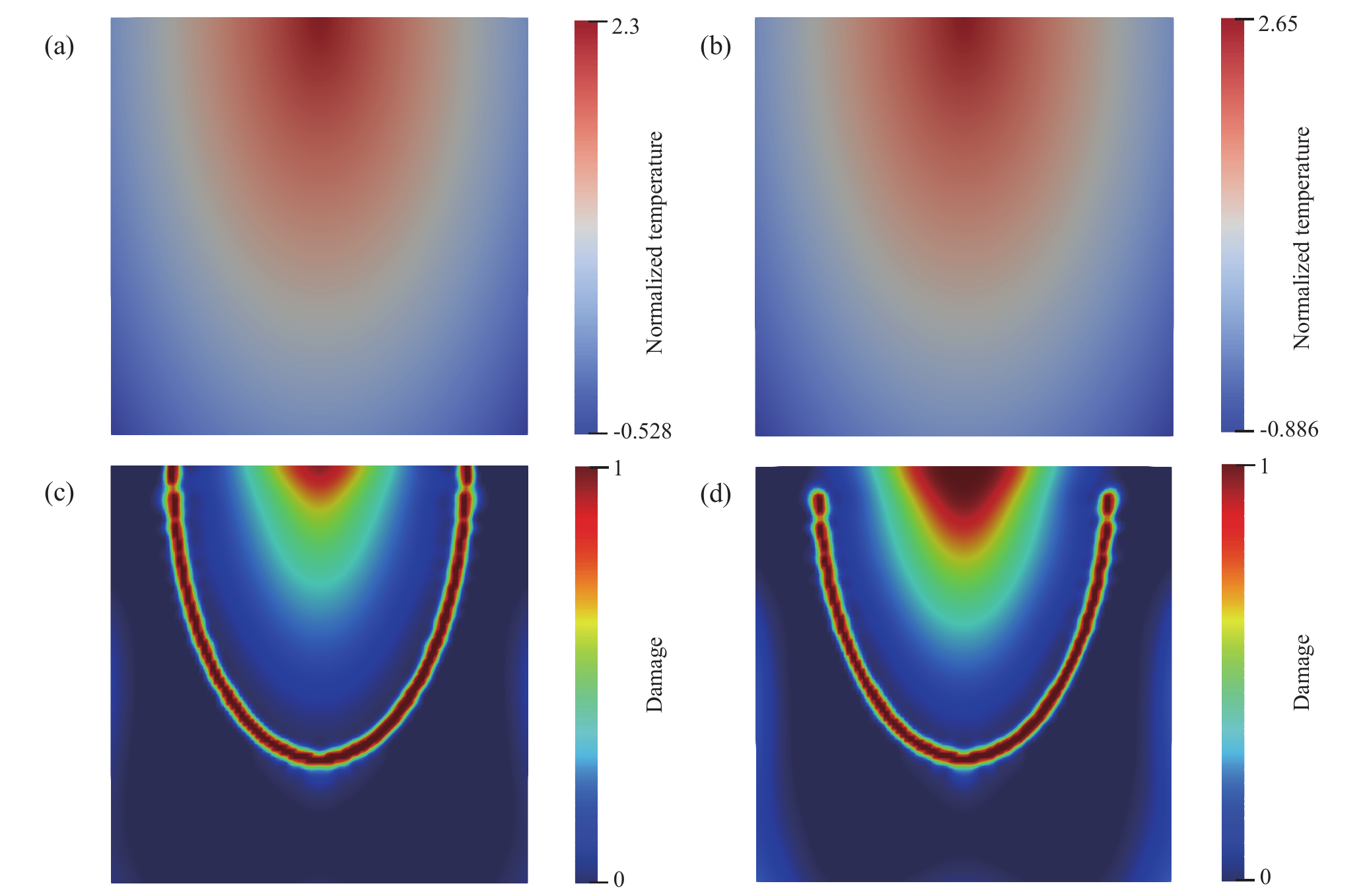}
    \caption{Normalized temperature profiles and hot cracking patterns of single track of PBF with different temperature gradient $c$. (a) $c=20$. (b) $c=25$.} 
    \label{fig:hot_crack_c}
\end{figure}

At this point, the result of keyhole mode differs considerably from the experimental result observed in the left column of Fig. \ref{fig:hot_crack_experiment}. Note that for this case apart from the circumferential crack, one observes a central crack as well. The disagreement of the crack pattern is the result of the way the temperature field is approximated, where the coefficient $c=10$ for all cases. In order to investigate the influence of the temperature gradient near the melting pool, further studies with different values for the parameter $c$ ($c=20,25$) are conducted. The approximated normalized temperature profiles and corresponding crack patterns are shown in Fig.~\ref{fig:hot_crack_c}. Compared with the results of the temperature field and crack pattern in Fig.~\ref{fig:hot_crack_T_d}(a) and (d) with $c=10$, it becomes clear that by increasing the parameter $c$, the temperature field has the similar distribution but the temperature gradient increases accordingly. As a result, the crack pattern also changes. The damage level near the center of the melting pool is getting bigger, meaning that apart from the circumferential crack, a central crack begins to show. However, because of the simple linear interpolation of temperature, the central region has the same relatively larger temperature gradient in all directions, which contributes to crack showing in a very diffusive region. Again, the latter observation does not fully agree with the experimental results in Fig.~\ref{fig:hot_crack_experiment}. This implies the necessity of having a more accurate temperature profile, which will be considered in the next subsection.


\subsubsection{Hot cracking simulation by using the phenomenological PBF thermal profile}
The analytical thermal profile has limited accuracy and does not allow parameter study on process condition. In this section, the numerical thermal profile calculated from the phenomenological thermal PBF model developed by the authors~\cite{yang2022validated, yi2019computational} is utilized as the temperature initial distribution before it cools down to room temperature. 
In the phenomenological thermal PBF model, the effective thermal properties of the powder bed and the resolidified phase are regarded explicitly. A phase indicator $\phi$ is introduced to indicate the state of the material, i.e. $\phi=1$ the fused state and $\phi=0$ the power bed. Readers are referred to\cite{yang2022validated, yi2019computational} for more details. Consider the phase-dependent thermal properties and the beam energy deposition, the heat transient problem is solved in the finite element method.

\begin{figure}[H]
    \centering
    \includegraphics[width=1.0\textwidth]{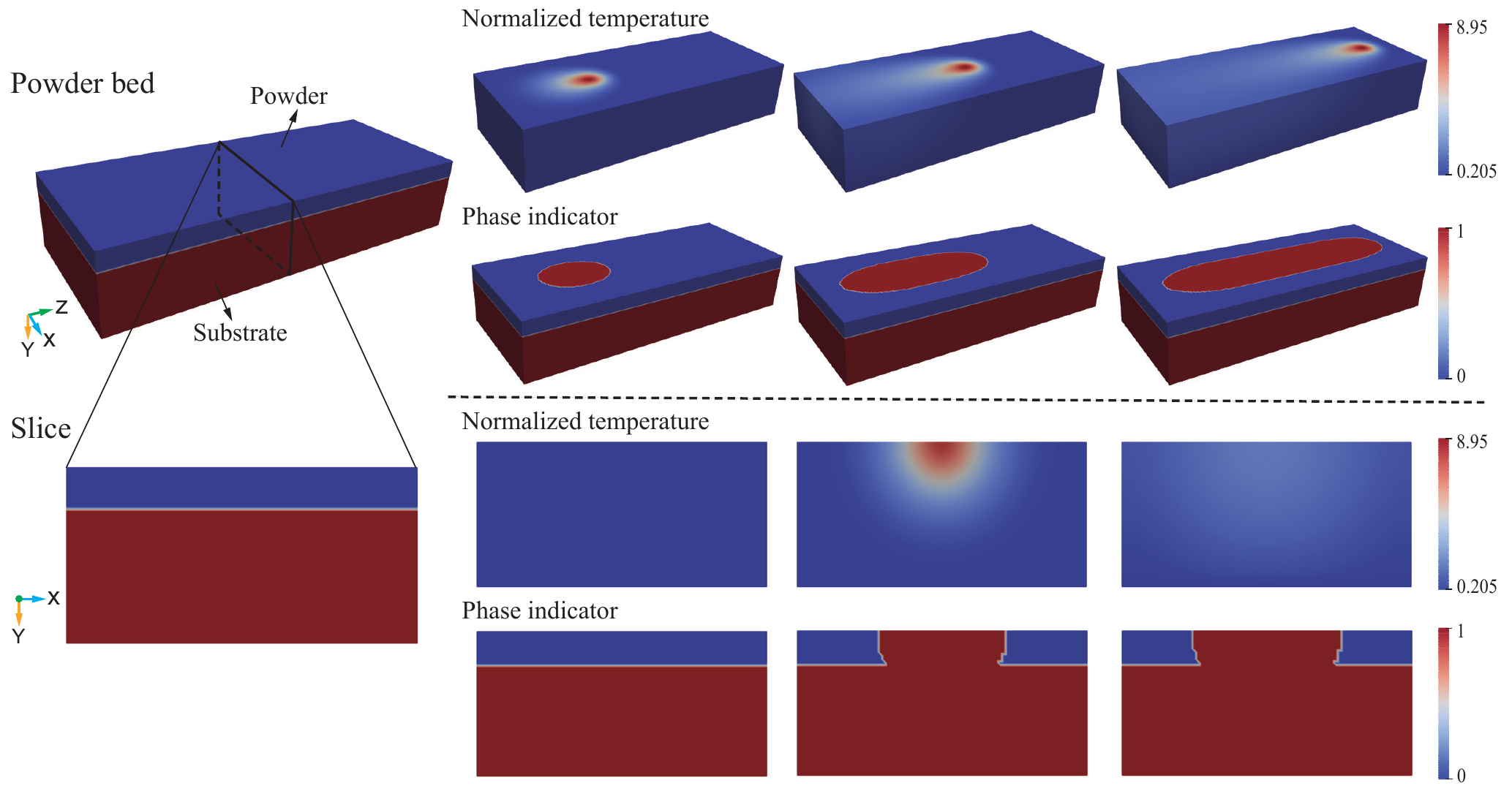}
    \caption{Temperature field and phase indicator evolution of the 3D phenomenological thermal PBF model (top) and the cross section perpendicular to the scanning direction (bottom).} 
    \label{fig:PBF}
\end{figure}


In this simulation, the domain has a volume of 1000 mm$\times$400 mm$\times$200 mm, with a 50-mm-thick powder bed layer and a 150-mm-thick substrate layer made of the same materials as the powder, as shown in Fig.~\ref{fig:PBF}. All the outer surfaces except the bottom surface of the powder bed are subject to the convection and radiation boundary conditions, while the bottom surface is applied with the Dirichlet boundary condition, i.e. the temperature is fixed. Meanwhile, the displacements in $x$, $y$ and $z$ direction of the bottom surface are fixed while other surfaces are traction-free. The laser power is $P_0$ = 2.18e4 W and the scanning speed is $v_0$ = 0.181 m/s. The other parameters used for this part are referred to the work \cite{yang2022validated}. 

In Fig.~\ref{fig:PBF}, the snapshots of the temperature field and the phase indicator evolution of the 3D phenomenological PBF model are shown. As the heat source like a laser beam moves forward ($z$), the temperature near the beam increases to the melting temperature, and the powders melt and solidify to the substrate. This process is reflected by the phase indicator changing from blue (0) to red (1). The snapshots of the temperature field and the phase indicator evolution of the slice along the scanning direction are also shown in Fig. \ref{fig:PBF}. As the laser approaches this slice, the temperature goes up and the powders melt. And as the beam passes by, the temperature drops while the materials keep as fused state.

After obtaining the steady-state cross-sectional thermal profile and the phase indicator distribution around the melt pool perpendicular to the scanning direction ($z$), they are transferred as the initial conditions for the subsequent hot cracking simulation. In the hot cracking simulation, a plane strain 2D case is assumed as the cross section is more or less mechanically constraint along the scanning direction. The bottom edge is fixed in the vertical direction ($y$) while the other edges are assumed to be traction-free. Similar as in the previous subsection, the solidification shrinkage is also considered. 

\begin{figure}[H]
    \centering
    \includegraphics[width=1.0\textwidth]{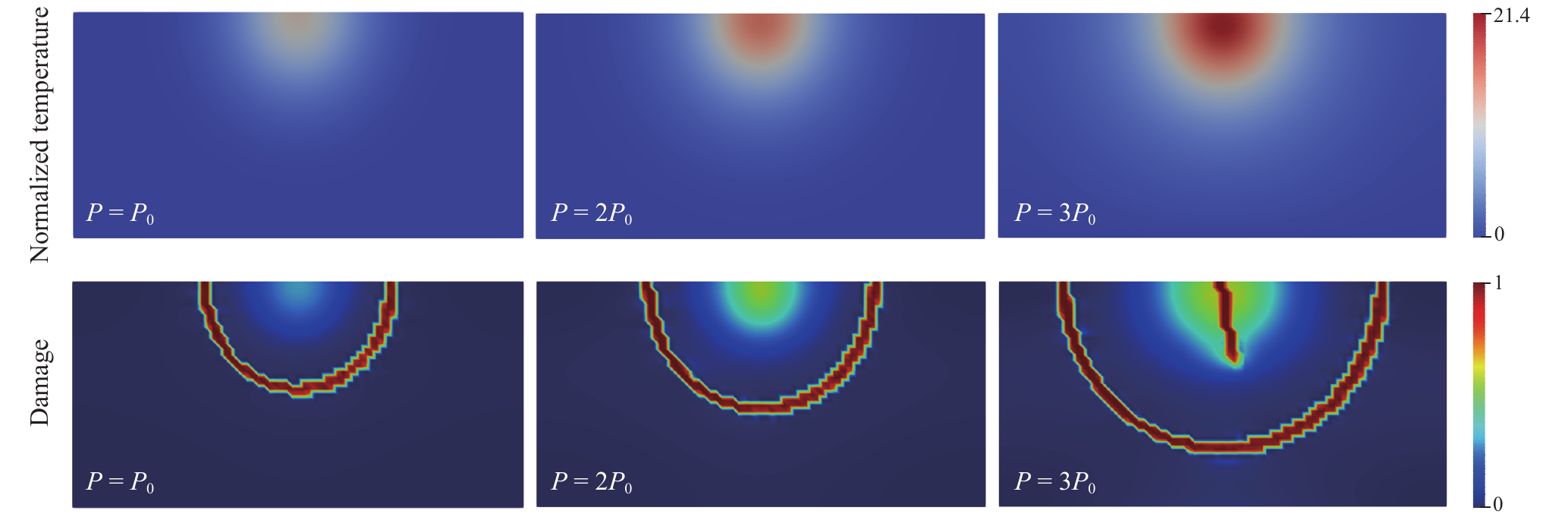}
    \caption{Hot cracking patterns of slice of PBF with different laser power.} 
    \label{fig:hot_crack_slice}
\end{figure}

\begin{figure}[H]
    \centering
    \includegraphics[width=1.0\textwidth]{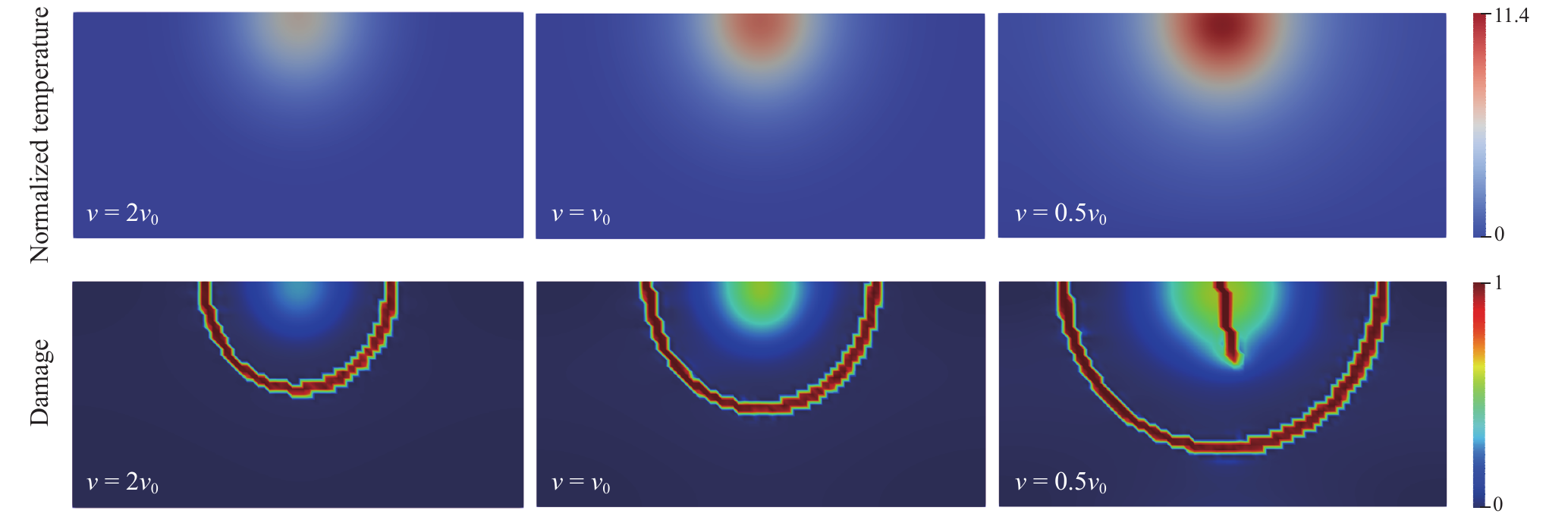}
    \caption{Hot cracking patterns of slice of PBF with different scanning speed.} 
    \label{fig:hot_crack_slice_speed}
\end{figure}

In order to study the influence of the process parameters like laser power ($P$) and scanning speed ($v$) on the hot cracking pattern, some parameter analysis are conducted.
Fig.~\ref{fig:hot_crack_slice} shows the different temperature profiles and the corresponding hot cracking patterns with varying laser power $P$ and fixed $v_0$. When the laser power is low ($P=P_0$), with the solidification shrinkage considered in the current model, only circumferential crack forms in the melting pool. In addition, a relatively low level of damage is also observed in the central region. Interestingly, this conduction mode hot cracking pattern predicted by numerical simulation is conformed in the results of the experimental observation of the conduction mode hot cracking showing in Fig. \ref{fig:hot_crack_experiment}. With the the laser power is increased to $2P_0$, the damage level of the central region is also increasing. Meanwhile, the circumferential crack expands outwards. The reason is the maximum and overall temperature of the melting pool increases, the temperature interval of the solidus temperature and liquidus temperature gets far from the center. When the laser power is further increased ($P=3P_0$), the damage level in the central region becomes even larger, and finally the central crack also arises, shifting the hot cracking pattern from the conduction mode to keyhole mode. 

The influence of the scanning speed $v$ is also studied, with fixed $P_0$, results of which are depicted in Fig.~\ref{fig:hot_crack_slice_speed}. As the scanning speed decreases from $v_0$ to $0.5v_0$, more energy inputs into the domain, causing the temperature and its gradient to increase. Therefore, the damage level in the central region is enhanced, and a central crack also appears, apart from the expanded circumferential crack, leading to the keyhole mode hot cracking. However, the hot cracking pattern tends to be the conduction mode when the scanning speed increases. The reason is the materials absorb less energy with fast scanning speed. Accordingly, the temperature and its gradient is reduced, and the damage level in the central region is also decreased. Thus, only the contracted circumferential crack forms in the melting pool.

The reasons for different hot cracking patterns under different process parameters are related to energy deposition and thus the thermal gradients. The laser power and the scanning speed determine the linear energy density $E_L=P/v$, which further determines the input energy to the PBF system. With larger laser power and slower scanning speed, $E_L$ becomes larger, and the temperature gradient near the melting pool is enhanced as well. Therefore, a larger thermal strain is formed and drives the crack to develop. The hot cracking pattern tends to be the keyhole mode. On the contrary, with smaller laser power and quicker scanning speed, $E_L$ becomes smaller, and the temperature gradient is low. Correspondingly, the hot cracking pattern tends to be the conduction mode.  These results are also in agreement with previous discussions. The comparisons between the numerical and experimental results above demonstrate the currently proposed modeling strategy has the potential to predict the hot cracking in the PBF and other AM processes. This study also provides some basic instructions on AM practice to eliminate hot cracking of AM products.  

\section{Conclusions and future work}
In this work, a thermodynamically consistent model for thermo-mechanically coupled phase-field fracture is introduced. Derived from the basic principles of thermodynamics, the coupling effects between mechanics, heat transfer, and fracture are all taken into account. Particularly, the influence of temperature-dependent fracture properties and the degradation of thermal conductivity with the crack field are studied to capture the temperature field and the crack pattern more accurately. The degraded thermal conductivity can avoid nonphysical heat transfer in the fully-cracked regions. The insensitivity of the CPF model, which is adopted in the multiphysics framework, with respect to the incorporated length scale parameter is also studied.

The proposed model is first applied to the canonical thermal fracture problem, i.e. the single-edge notched tension. The length scale insensitive of CPF in multiphysics is compared with AT2 model, and the degradation of thermal conductivity with crack is also compared with constant property. 
The model is further validated by the quenching test. The crack mechanism like crack initiation, propagation, stop and further propagation with different initial temperatures follows the same pattern. However, the crack pattern still keeps evolving with the increase of the initial temperature. The numerical predictions are validated by the experimental results.

The model is further applied to study the hot cracking phenomenon in AM, particularly PBF. The analytic temperature solution is first used to obtain the temperature field near the melting pool. When the solidification shrinkage is considered, the predicted hot cracking patterns of the conduction mode AM are mainly shown as the circumferential crack. While for the keyhole mode AM, the circumferential crack and central crack shows simultaneously in the melting pool. Subsequently, the phenomenological PBF model is utilized to obtain a more accurate temperature field, specifically for a slice of the whole powder bed model. The process parameters like the laser power and the scanning speed on the final crack pattern are investigated. The results show that a higher laser power and slower scanning speed is favorable of keyhole mode hot cracking while a lower laser power and quicker scanning speed result in the conduction mode cracking pattern. The comparisons of the hot crackings between the numerical results and the experimental observations shows good agreement, demonstrating the capability of the proposed model for the prediction of hot cracking in PBF and other AM processes and the potential for further studies.

The current investigations open up new possibilities for studies on the hot cracking of AM in a numerical way. Further study can be performed when it comes to a more precise thermal fracture model. In this work, the elastic and brittle fracture is used, while in future work the influence of plasticity and extension to ductile fracture would also be of great interest. Using a more accurate temperature profile and phase evolution of the material is also interesting to study. For now, the analytical temperature solution and the phenomenological phase-field model of the PBF process are used to obtain the thermal field. However, a more high-fidelity powder-level PBF model can be used to track the temperature evolution history and powder evolution history in future work, as it is shown in our related work on selective sintering \cite{yang20193d, zhou20213d}. The simulation results above are going to be combined with the thermal fracture model to capture the hot cracking behaviors of the powder-level PBF model more accurately. Finally, other process parameters except for laser power and scanning speed in the PBF process are also worthy of investigation, so that the potential process window for PBF to eliminate hot cracking can be proposed.

\section*{Declaration of interests}
The authors declare that they have no known competing financial interests or personal relationships that could have appeared to influence the work reported in this paper.

\section*{Acknowledgments}
The authors gratefully acknowledge the financial support from the China Scholarship Council (CSC). The authors also greatly appreciate their access to the Lichtenberg High-Performance Computer and the technique supports from the HHLR, Technische Universität Darmstadt.

\newpage
\appendix
\section{Derivation of the dissipation inequality}
\subsection{Energy conservation: first principle}
The first law of thermodynamics represents a detailed balance describing the interplay between the internal energy, the kinetic energy, the rate at which power is expended, and the heat transferred to $\Omega$, which is expressed by:
\begin{equation} \label{eq:1}
    \dfrac{d}{dt}(\mathcal{E}+\mathcal{K})=\mathcal{W}+\mathcal{Q},
\end{equation}
where $\mathcal{E}$,  $\mathcal{K}$, $\mathcal{W}$ and $\mathcal{Q}$ denote the net internal energy, the kinetic energy, the external power, and the heat flow, respectively. 
\begin{equation} \label{eq:2}
    \left\{\begin{array}{ll}
&\mathcal{E}=\int_{\Omega} \rho e d V\\
&\mathcal{K} =\int_{\Omega} \dfrac{1}{2} \rho\boldsymbol {\dot x}^{2} d V\\
&\mathcal{W}=\int_{\partial \Omega} \boldsymbol T \cdot \boldsymbol {\dot x} d S+\int_{\Omega} \boldsymbol{b} \cdot \boldsymbol{{\dot x}} d V\\
&\mathcal{Q}=-\int_{\partial \Omega} \boldsymbol{q} \cdot \boldsymbol{n} dS+\int_{\Omega} Q d V
\end{array}\right.
\end{equation}
where $\rho$ denotes the material density, $e$ internal energy per unit mass, $(\boldsymbol {\dot x})$ is the time derivative of displacement, $\boldsymbol T$ denotes the surface traction, and $\boldsymbol b$ the generalized body force, $\boldsymbol q$ is the heat flux, $\boldsymbol n$ the outward unit normal to $\partial \Omega$ and $Q$ internal heat supply.

Substituting the equations in Eq. \ref{eq:2} into energy balance Eq. \ref{eq:1} yields
\begin{equation} \label{eq:3}
    \dfrac{d}{dt}{\int_{\Omega} \rho\left(e+\dfrac{1}{2}\boldsymbol{{\dot x}}^{2}\right) d V}=-\int_{\partial \Omega} \boldsymbol{q} \cdot \boldsymbol{n} dS+\int_{\Omega} Q dV+\int_{\partial \Omega} \boldsymbol{T}  \cdot \boldsymbol{{\dot x}} dS+\int_{\Omega} \boldsymbol{b} \cdot \boldsymbol{{\dot x}} dV.\\
\end{equation}

By integrating the balance equation of linear momentum (i.e. Newton's law), multiplied by the derivative of displacement over the domain, one can get
\begin{equation} \label{eq:4}
    -\int_{\Omega} \boldsymbol \sigma: \dot{\boldsymbol\varepsilon}dV+\int_{\partial \Omega} \boldsymbol{T}  \cdot \boldsymbol{{\dot x}} d S+\int_{\Omega} \boldsymbol{b} \cdot \boldsymbol{{\dot x}} d V=\dfrac{d }{dt}\int_{\Omega} \dfrac{1}{2}\boldsymbol{{\dot x}}^{2}dV.
\end{equation}
Substituting Eq. \ref{eq:4} into Eq. \ref{eq:3} and applying the divergence theorem, we obtain the following energy equilibrium equation
\begin{equation}\label{eq:5}
    \int_{\Omega}(\rho \dot{e}-\boldsymbol \sigma: \dot{\boldsymbol\varepsilon}+\nabla\cdot \boldsymbol{q}+Q) d V=0.
\end{equation}

The identity in Eq. \ref{eq:5} is valid for any region, thus the local energy balance equation is obtained
\begin{equation}\label{eq:6}
    \rho \dot{e}=\boldsymbol \sigma: \dot{\boldsymbol\varepsilon}-\nabla\cdot \boldsymbol{q}+Q.
\end{equation}

In thermo-mechanical phase-field fracture analysis, besides the temperature $T$ and the total strain $\boldsymbol\varepsilon$, the state variables also include the phase=field crack $d$. Correspondingly, the external power $\mathcal{W}$ is reformulated as
\begin{equation}\label{eq:7}
    \mathcal{W}=\int_{\partial \Omega}( \boldsymbol T \cdot \boldsymbol {\dot x} +\boldsymbol H\cdot \nabla\dot d)d S+\int_{\Omega} (\boldsymbol{b} \cdot \boldsymbol{{\dot x}} +K\dot d )d V,
\end{equation}
where $ H$ is micro-traction on crack surfaces and $K$ denotes the internal micro-forces, which will be discussed later in the following derivation. Thus the energy balance equation, which accounts for thermal diffusion and power produced by the micro and macro forces, is given as ( Stumpf and Hackl, 2003 )
\begin{equation}\label{eq:8}
    \rho \dot{e}=\boldsymbol\sigma: \dot{\boldsymbol\varepsilon}+K \dot{d}+\boldsymbol H \cdot \nabla \dot{d}-\nabla \cdot \boldsymbol{q}+Q.
\end{equation}

\subsection{ Entropy inequality: second principle}
The second law of thermodynamics postulates that the rate of net entropy production $\dot s$ in any convecting spatial region $\Omega$ is always nonnegative
\begin{equation}\label{eq:9}
    \dot s =\int_{\Omega}\rho \dot \eta dV-(-\int_{\partial \Omega} \dfrac{\boldsymbol q}{T}\cdot \boldsymbol n+\int_{\Omega}\dfrac{Q}{T})dV\geq 0,
\end{equation}
where $\eta$ is the specific entropy per unit mass, the first term on the right hand denotes the internal entropy and the second term denotes entropy flow, which is the rate at which entropy is transferred to $\Omega$. 

This inequality is valid for any region of the body and using the divergence theorem leads to the following local form of the irreversibility of the entropy production rate:
\begin{equation}\label{eq:10}
    \rho \dot \eta+ \nabla\cdot (\dfrac{\boldsymbol q}{T})-\dfrac{Q}{T} \geq0.
\end{equation}
The fundamental inequality containing the first and second principles is obtained by replacing $Q $ with the expression resulting from Eq. \ref{eq:8} of conservation of energy:
\begin{equation}\label{eq:11}
    \rho  \dot{\eta}+ \nabla \cdot(\dfrac{\boldsymbol{q}}{T})-\dfrac{1}{T}\left(\rho \dot e-\boldsymbol{\sigma}: \boldsymbol{\varepsilon}-K \dot{d}-\boldsymbol H \cdot \nabla \dot{d}+\nabla\cdot\boldsymbol{q}\right) \geqslant 0.
\end{equation}
Note that
\begin{equation}\label{eq:12}
    \nabla\cdot (\dfrac{\boldsymbol{q}}{T})=\dfrac{1}{T}\nabla\cdot \boldsymbol{q}-\dfrac{1}{T^2} \nabla T \cdot \boldsymbol{q},
\end{equation}
we obtain
\begin{equation}\label{eq:13}
    \rho(T \dot{\eta}-\dot e)+\boldsymbol \sigma: \boldsymbol{\varepsilon}+K \dot{d}+\boldsymbol H \cdot \nabla \dot{d}-\dfrac{1}{T} \nabla T \cdot \boldsymbol{q} \geq 0.
\end{equation}
Here we introduce the specific free energy $\psi$ defined by the Legendre transforms (shell, 2015)
\begin{equation}\label{eq:14}
    \psi=e-T \eta,
\end{equation}
where $\psi$ is the Helmholtz free energy per unit mass, which measures the amount of obtainable work in a closed thermodynamic system. The rate form of the internal energy $e$ is given by
\begin{equation}\label{eq:15}
    \dot{\psi}=\dot{e}-\dot{T} \eta-T \dot{\eta}.
\end{equation}
Substituting Eq. \ref{eq:15} into Eq. \ref{eq:13}, the internal energy can be eliminated from the energy balance equation, and the local Clausius-Duhem inequality is obtained.

\newpage
\bibliographystyle{elsarticle-num}


\end{document}


\begin{frontmatter}

\appendix A
\subsubsection{ Energy conservation: first principle}
The first law of thermodynamics represents a detailed balance describing the interplay between the internal energy, the kinetic energy, the rate at which power is expended and the heat transferred to $\Omega$, which is expressed by:
\begin{equation} \label{eq:1}
    \frac{d}{dt}(\mathcal{E}+\mathcal{K})=\mathcal{W}+\mathcal{Q}
\end{equation}
where $\mathcal{E}$,  $\mathcal{K}$, $\mathcal{W}$ and $\mathcal{Q}$ denote the net internal energy, the kinetic energy, the external power and the heat flow, respectively. 
\begin{equation} \label{eq:2}
    \left\{\begin{array}{ll}
&\mathcal{E}=\int_{\Omega} \rho e d V\\
&\mathcal{K} =\int_{\Omega} \frac{1}{2} \rho\pmb {\dot x}^{2} d V\\
&\mathcal{W}=\int_{\partial \Omega} \pmb T \cdot \pmb {\dot x} d S+\int_{\Omega} \pmb{b} \cdot \pmb{{\dot x}} d V\\
&\mathcal{Q}=-\int_{\partial \Omega} \pmb{q} \cdot \pmb{n} dS+\int_{\Omega} Q d V
\end{array}\right.
\end{equation}
where $\rho$ denotes the material density, $e$ internal energy per unit mass, $(\pmb {\dot x})$ is the time derivative, $\pmb T$ denotes the surface traction and $\pmb b$ the generalized body force, $\pmb q$ is the heat flux, $\pmb n$ the outward unit normal to $\partial \Omega$ and $Q$ internal heat supply.
Substituting the equations in Eq.\ref{eq:2} into energy balance Eq.\ref{eq:1} yields
\begin{equation} \label{eq:3}
    \frac{d}{dt}{\int_{\Omega} \rho\left(e+\frac{1}{2}\pmb{{\dot x}}^{2}\right) d V}=-\int_{\partial \Omega} \pmb{q} \cdot \pmb{n} dS+\int_{\Omega} Q dV+\int_{\partial \Omega} \pmb{T}  \cdot \pmb{{\dot x}} dS+\int_{\Omega} \pmb{b} \cdot \pmb{{\dot x}} dV\\
\end{equation}

By applying the axiom of equilibrium of the principle of virtual power to the region, for any virtual movement, the virtual power of the acceleration quantities (i.e., inertia forces) is equal to the sum of the virtual power of internal forces and of external forces, which is expressed as
\begin{equation} \label{eq:4}
    -\int_{\Omega} \pmb \sigma: \dot{\pmb\varepsilon}dV+\int_{\partial \Omega} \pmb{T}  \cdot \pmb{{\dot x}} d S+\int_{\Omega} \pmb{b} \cdot \pmb{{\dot x}} d V=\frac{d }{dt}\int_{\Omega} \frac{1}{2}\pmb{{\dot x}}^{2}dV
\end{equation}
Substituting Eq.\ref{eq:4} into Eq.\ref{eq:3} and applying the divergence theorem, we obtain the following equilibrium equations
\begin{equation}\label{eq:5}
    \int_{\Omega}(\rho \dot{e}-\pmb \sigma: \dot{\pmb\varepsilon}+\nabla\cdot \pmb{q}+Q) d V=0
\end{equation}

The identity in Eq.\ref{eq:5}is valid for any region, thus the local energy balance equation is obtained
\begin{equation}\label{eq:6}
    \rho \dot{e}=\pmb \sigma: \dot{\pmb\varepsilon}-\nabla\cdot \pmb{q}+Q
\end{equation}

In thermo-mechanical phase field fracture analysis, besides the temperature $T$ and the total strain $\pmb\varepsilon$, the state variables also includes the crack phase field $d$. Correspondingly, the external power $\mathcal{W}$ is reformulated as
\begin{equation}\label{eq:7}
    \mathcal{W}=\int_{\partial \Omega}( \pmb T \cdot \pmb {\dot x} +\pmb H\cdot \nabla\dot d)d S+\int_{\Omega} (\pmb{b} \cdot \pmb{{\dot x}} +K\dot d )d V
\end{equation}
where $ H$ is micro-traction on crack surfaces and $K$ denotes the internal micro-forces, which will be discussed later in the following derivation. Thus the energy balance equation, which accounts for thermal diffusion and power produced by the micro and macro forces, is given as ( Stumpf and Hackl, 2003 )
\begin{equation}\label{eq:8}
    \rho \dot{e}=\pmb\sigma: \dot{\pmb\varepsilon}+K \dot{d}+\pmb H \cdot \nabla \dot{d}-\nabla \cdot \boldsymbol{q}+Q
\end{equation}

\subsubsection{ Entropy inequality: second principle}
The second law of thermodynamics postulates that the rate of net entropy production $\dot s$ in any convecting spatial region $\Omega$ is always nonnegative
\begin{equation}\label{eq:9}
    \dot s =\int_{\Omega}\rho \dot \eta dV-(-\int_{\partial \Omega} \frac{\pmb q}{T}\cdot \pmb n+\int_{\Omega}\frac{Q}{T})dV\geq 0
\end{equation}
where $\eta$ is the specific entropy per unit mass, the first term on the right hand denotes the internal entropy and the second term denotes entropy flow, which is the rate at which entropy is transferred to $\Omega$. 
This inequality is valid for any region of the body and using the divergence theorem leads to the following local form of the irreversibility of the entropy production rate:
\begin{equation}\label{eq:10}
    \rho \dot \eta+ \nabla\cdot (\frac{\pmb q}{T})-\frac{Q}{T} \geq0
\end{equation}
The fundamental inequality containing the first and second principles is obtained by replacing $Q $ with the expression resulting from Eq. (8) of conservation of energy:
\begin{equation}\label{eq:11}
    \rho  \dot{\eta}+ \nabla \cdot(\frac{\boldsymbol{q}}{T})-\frac{1}{T}\left(\rho \dot e-\boldsymbol{\sigma}: \pmb{\varepsilon}-K \dot{d}-\pmb H \cdot \nabla \dot{d}+\nabla\cdot\pmb{q}\right) \geqslant 0
\end{equation}
Note that
\begin{equation}\label{eq:12}
    \nabla\cdot (\frac{\pmb{q}}{T})=\frac{1}{T}\nabla\cdot \pmb{q}-\frac{1}{T^2} \nabla T \cdot \boldsymbol{q}
\end{equation}
we obtain
\begin{equation}\label{eq:13}
    \rho(T \dot{\eta}-\dot e)+\pmb \sigma: \pmb{\varepsilon}+K \dot{d}+\pmb H \cdot \nabla \dot{d}-\frac{1}{T} \nabla T \cdot \boldsymbol{q} \geq 0
\end{equation}
Here we introduce the specific free energy $\psi$ defined by the Legendre transforms (shell, 2015)
\begin{equation}\label{eq:14}
    \psi=e-T \eta
\end{equation}
where $\psi$ is the Helmholtz free energy per unit mass, which measures the amount of obtainable work in a closed thermo-dynamic system. The rate form of the internal energy $e$ is given by
\begin{equation}\label{eq:15}
    \dot{\psi}=\dot{e}-\dot{T} \eta-T \dot{\eta}
\end{equation}
Substituting Eq.\ref{eq:15} into Eq.\ref{eq:13}, the internal energy can be eliminated from the energy balance equation, and the local Clausius-Duhem inequality is obtained
\\
\\
\\

In this work,  the independent field variables are $(\pmb u,d,T)$, which must be determined as solution of the problem. The rest unknown field variables need to be given by constitutive equations in terms of the independent field variables. For the thermomechanical damage problem, we assume the constitutive equations in the general form (H. Stumpf, K. Hackl, 2003)
\begin{equation}\label{eq:17}
    \psi=\psi(\pmb\varepsilon,d,\nabla d,T, \dot{\pmb\varepsilon},\dot d,\Delta d,\nabla T)
\end{equation}
The damage constitutive equations must be consistent with the balance law and the local Clausius-Duhem inequality. Introducing the constitutive equations Eq.\ref{eq:17} into Eq.\ref{eq:16} yields

\begin{equation}
\begin{aligned}
\mathcal D&=(\pmb\sigma-\rho \frac{\partial \psi}{\partial \pmb\varepsilon}): \dot{\pmb\varepsilon}+(K- \rho \frac{\partial \psi}{\partial d})\dot{d}+ (\pmb H-\rho \frac{\partial \psi}{\partial \nabla d}) \cdot \nabla \dot{d}-(\eta+\rho \frac{\partial \psi}{\partial T})\dot T \\
&-\rho \frac{\partial \psi}{\partial \dot{\pmb\varepsilon}}\ddot{ \pmb\varepsilon}-\rho \frac{\partial \psi}{\partial \dot d}\ddot d-\rho \frac{\partial \psi}{\partial \nabla \dot d}\nabla \ddot d-\rho \frac{\partial \psi}{\partial \nabla T}\nabla \dot T-\frac{1}{T} \nabla T \cdot \boldsymbol{q} \geq 0
\end{aligned}
\end{equation}
The inequality must be satisfied identically by all constitutive equations, thus it follows 
\begin{equation}\label{eq:19}
    \frac{\partial \psi}{\partial \dot{\pmb\varepsilon}}=0,\quad \frac{\partial \psi}{\partial \dot d}=0,\quad
\frac{\partial \psi}{\partial \nabla \dot d}=0,\quad
\frac{\partial \psi}{\partial \nabla T}=0
\end{equation}
Hence, the Helmholtz free energy $\psi$ depends neither on $(\dot{\pmb\varepsilon},\dot d,\Delta d)$ nor $\nabla T$, so $\psi$ is a function of $(\pmb\varepsilon,  d, \nabla d,T)$.

\appendix B

2.the volumetric-deviatoric split, which is based on the decomposition of the strain tensor into spherical and deviatoric components.
\begin{equation}\label{eq:35}
    \pmb\varepsilon_e=\pmb\varepsilon_{S}+\pmb\varepsilon_{D}, \quad \pmb\varepsilon_{S}=\frac{1}{n} {tr}(\pmb\varepsilon) I, \quad \pmb\varepsilon_{D}=\pmb\varepsilon_e-\frac{1}{n} {tr}(\pmb\varepsilon) I
\end{equation}
where $I$ denotes the n-dimensional identity tensor. With this decomposition, the strain energy density of a linear elastic isotropic material may be written as the sum of the spherical and deviatoric contributions:
\begin{equation}\label{eq:36}
    \psi_e=\frac{1}{2} \lambda ({tr}\pmb\varepsilon_e)^{2}+\mu \pmb\varepsilon_e \cdot \pmb\varepsilon_e=\kappa_0\frac{({tr}\pmb\varepsilon_e)^{2}}{2}+\mu \pmb\varepsilon_D \cdot \pmb\varepsilon_D
\end{equation}
where $\kappa_0=\lambda+\frac{2\mu}{n}$  is the bulk modulus of the material.

We introduce the decomposition of the trace of the strain tensor in positive parts: ${\left\langle tr \pmb\varepsilon_e\right\rangle^{+}}=tr\langle\pmb\varepsilon_e\rangle$ and negative parts: $\langle{tr}\pmb\varepsilon_e\rangle^{-}=tr\langle-\pmb\varepsilon_e\rangle$. The Macaulay bracket operator is defined as
\begin{equation}
    \langle x\rangle=\left\{\begin{array}{ll}
x & \text { if } x \geq 0 \\
0 & \text { if } x<0
\end{array}\right.
\end{equation}
and further distinguish the contributions due to compression, expansion, and shear of the strain energy as follows:
\begin{equation}
    \psi_e=\kappa_{0} \frac{({\left\langle tr \pmb\varepsilon_e\right\rangle^{-}})^{2}}{2}+\kappa_{0} \frac{({\left\langle tr \pmb\varepsilon_e\right\rangle^{+}})^{2}}{2}+\mu\pmb \varepsilon_{D} \cdot \pmb\varepsilon_{D}
\end{equation}
In this split, tensile volumetric and deviatoric strains contribute to damage. 
\begin{equation}
    \psi_e^+=\kappa_{0} \frac{({\left\langle tr\pmb \varepsilon_e\right\rangle^{+}})^{2}}{2}+\mu {\pmb\varepsilon}_{D} \cdot{\pmb\varepsilon}_{D}
\end{equation}

\begin{equation}
    \psi_e^-=\kappa_{0} \frac{({\left\langle tr\pmb \varepsilon_e\right\rangle^{-}})^{2}}{2}
\end{equation}
The stress tensor can be derived as
\begin{equation}
    \pmb \sigma=\frac{\partial \psi_e}{\partial \pmb \varepsilon}=g(d)\frac{\partial \psi_e^+}{\partial \pmb \varepsilon} +\frac{\partial \psi_e^-}{\partial \pmb \varepsilon}=g(d)\pmb \sigma^+ +\pmb \sigma^-
\end{equation}
where
\begin{equation}
    \pmb \sigma^+ =\kappa_{0} \langle{tr}\pmb\varepsilon_e\rangle^{+}\pmb I+2\mu\pmb \varepsilon_{D}, \quad
\pmb \sigma^-=\kappa_{0} \langle{tr}\pmb\varepsilon_e\rangle^{-}\pmb I
\end{equation}
